\documentclass[10pt,twoside,twocolumn,english,aip,pop]{revtex4-1}
\usepackage[T1]{fontenc}
\usepackage[latin9]{inputenc}
\usepackage{amsmath}
\usepackage{amssymb}
\usepackage{graphicx}
\usepackage{esint}

\makeatletter
 
 \@ifundefined{textcolor}{}
 {%
   \definecolor{BLACK}{gray}{0}
   \definecolor{WHITE}{gray}{1}
   \definecolor{RED}{rgb}{1,0,0}
   \definecolor{GREEN}{rgb}{0,1,0}
   \definecolor{BLUE}{rgb}{0,0,1}
   \definecolor{CYAN}{cmyk}{1,0,0,0}
   \definecolor{MAGENTA}{cmyk}{0,1,0,0}
   \definecolor{YELLOW}{cmyk}{0,0,1,0}
 }

\@ifundefined{showcaptionsetup}{}{%
 \PassOptionsToPackage{caption=false}{subfig}}
\usepackage{subfig}
\makeatother

\usepackage{babel}
\begin{document}

\title{Kinetic simulation of the O-X conversion process in dense magnetized
plasmas}

\author{M. Ali Asgarian}

\email{maliasgarian@ph.iut.ac.ir, maa@msu.edu}

\selectlanguage{english}%

\address{Physics Department, Isfahan University of Technology, Isfahan, Iran}

\address{Department of Electrical and Computer Engineering, Michigan State
University, Michigan, USA}

\author{J. P. Verboncoeur}

\address{Department of Electrical and Computer Engineering, Michigan State
University, Michigan, USA}

\author{A. Parvazian}

\address{Physics Department, Isfahan University of Technology, Isfahan, Iran}

\author{R. Trines}

\address{STFC Rutherford Appleton Laboratory, Didcot, United Kingdom}
\begin{abstract}
The ordinary-extraordinary-Bernstein (O-X-B) double conversion is
considered and the O-X conversion simulated with a kinetic particle
model for parameters of the TJ-II stellarator. This simulation has
been done with the particle-in-cell code, XOOPIC (X11-based object-oriented
particle-in-cell). XOOPIC is able to model the non-monotonic density
and magnetic profile of TJ-II. The first step of conversion, O-X conversion,
is observed clearly. By applying some optimizations, such as increasing
the grid resolution and number of computational particles in the region
of the X-B conversion, the simulation of the second step is also possible.
By considering the electric and magnetic components of launched and
reflected waves, the O-mode wave and the X-mode wave can be easily
detected. Via considering the power of the launched O-mode wave and
the converted X-mode wave, the efficiency of O-X conversion for the
best theoretical launch angle is obtained, which is in good agreement
with previous computed efficiencies via full-wave simulations. For
the optimum angle of $47^{\circ}$ between the wave-vector of the
incident O-mode wave and the external magnetic field, the conversion
efficiency is $66\%$.
\end{abstract}
\maketitle

\section{Introduction}

\subsection{Motivation}

For ensuring energy security and reduction of environmental problems,
it is necessary for the world to move toward a mix of power sources
such as fossil fuels, nuclear fission, fusion, and renewables. Presently,
fossil fuels are the primary power source, with acute problems in
environmental pollution and declining natural resources. Nuclear fission
poses problems of nuclear waste, proliferation, accidental radiation
release in coolant system breaches, and potentially catastrophic loss
of coolant accidents such as Fukishima and Chernobyl. The renewable
energy sources are dependent upon environmental conditions such as
sunlight, wind, or water flow, and therefore are not continuous nor
scalable. Fusion would be a source without the supply and environmental
problems of fossil fuels, the waste, proliferation, and safety problems
of fission, and the inability to supply consistent base load power
of renwables. In the near future, fusion can provide a virtually inexhaustible
source of energy \cite{energy security,Fukishima and Chernobyl,Environmental problems}.

Energy producing fusion reactions require substantial heating and
confinement. So far, the tokamak is the most promising magnetic confinement
device to approach conditions necessary for net energy production.
Heating the core of tokamak plasmas remains one of the key challenges
in achieving fusion temperatures. One method of heating is to excite
plasma waves by launching electromagnetic waves through the edge plasma
into core. Dissipation of the kinetic energy of the plasma wave motion
via collisions can result in heating of the plasma. Radio frequency
heating is typically in three frequency ranges: the electron gyro
frequency, the lower hybrid frequency, and the ion gyro frequency.
The launched electromagnetic wave propagates into the plasma until
it reaches a location where the magnetic field strength and the plasma
density are such that one of the plasma resonance frequencies equals
the wave frequency, at which point the energy of the external wave
is transferred totally or partially to the plasma to excite waves
in the plasma. Finally, the energy of the plasma waves is dissipated
by collisions among the particles, thereby heating the plasma. A special
kind of electron cyclotron (EC) wave, the electron Bernstein wave
(EBW), is useful for heating the plasma core since it can penetrate
beyond the cutoffs. The processes of EBW excitation have been used
and simulated. Here we simulate one of these processes using the particle
in cell model.

\subsection{Electron Bernstein Waves}

The electron Bernstein wave (EBW) is a special electrostatic cyclotron
wave which propagates with a short wavelength in a magnetized hot
plasma \cite{Laqua Review}. After improving microwave sources to
powers higher than $100\,\textrm{kW}$ and frequencies in the range
of $\sim100\,\textrm{GHz}$, and the possibility of providing over-dense
plasmas in devices like Spherical Tokamaks (STs) and Stellarators,
the practical use of EBW, now some years after theoretical discussions,
is possible. 

Due to the wave cut-off surfaces in the plasma, the electromagnetic
electron cyclotron waves, e.g. the Ordinary (O) and Extraordinary
(X) modes, cannot propagate in over-dense regions, so the EBW is used
for heating, driving current and doing temperature diagnostics beyond
these surfaces. The wavelength of the EBW is comparable with the electron
gyro radius (Larmor radius). To obtain the dispersion relation, we
must take into account the effects of the finite Larmor radius on
the propagating wave. The Larmor radius is strongly dependent on the
thermal velocity and magnetic field, so we should include the temperature
of plasma species and the effect of magnetic field on the direction
of wave propagation in the calculation of the dielectric tensor. In
this case, we use the hot dielectric tensor \cite{Stix Book}. In
this level of approximation, in addition to the O-mode and X-mode
electromagnetic waves obtained in the cold approximation, we have
a mode named the Bernstein wave \cite{Bernstein Waves}. The longitudinal
electrostatic EBW is perpendicular to the external magnetic field
so the generated electric field is also perpendicular to the magnetic
field, i.e. $\mathbf{k}\parallel\mathbf{E}\perp\mathbf{B_{\mathbf{0}}}$.

\begin{figure*}
\subfloat[\label{fig:Dispersion Curve of O and X Waves-1a}]{\includegraphics[scale=0.3]{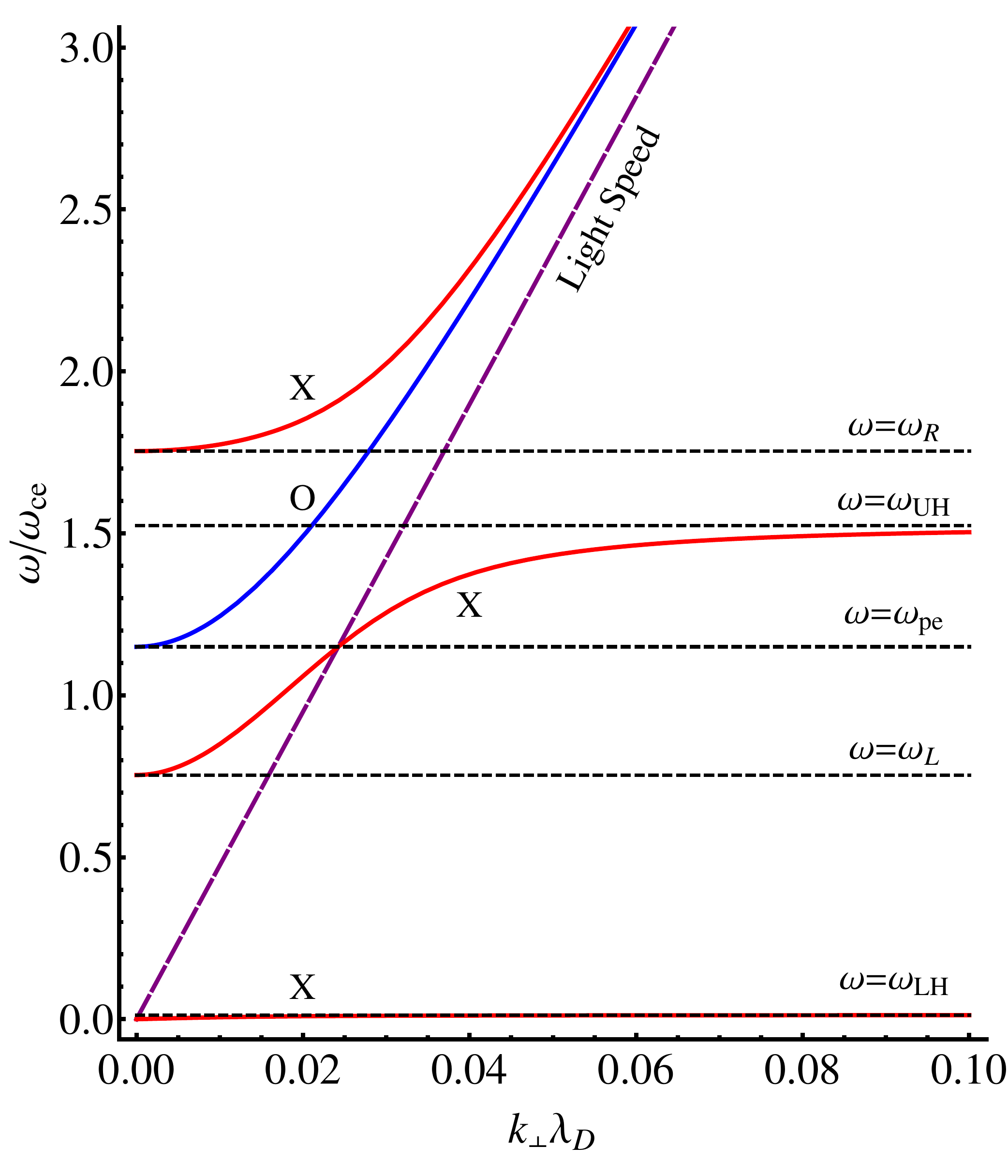}}\subfloat[\label{fig:Dispersion Curve of O and X Waves-1b}]{\includegraphics[scale=0.3]{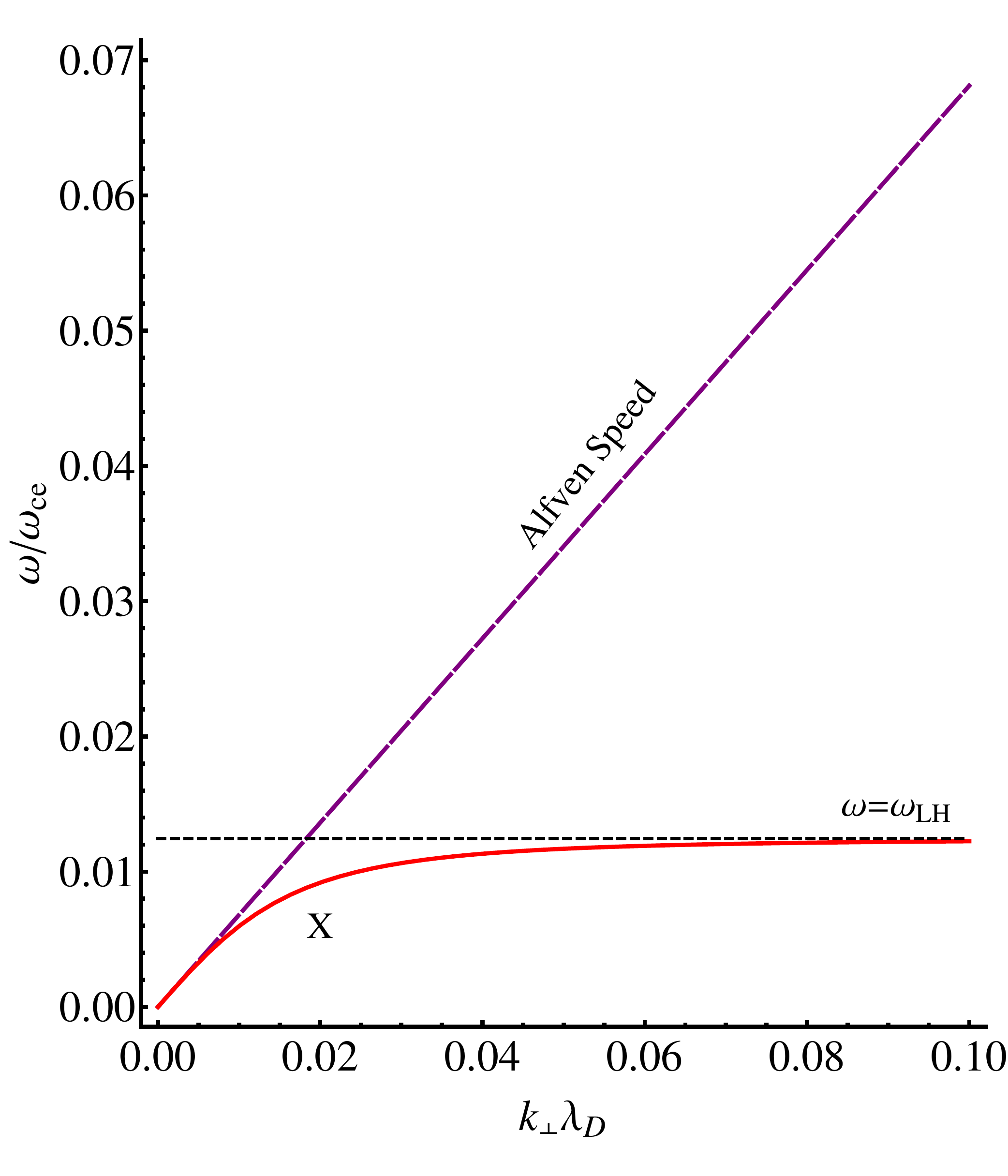}}\caption{(a)The dispersion curve of O-mode and X-mode waves. (b) The detailed
shape of lower branch of the X-mode wave. \label{fig:Dispersion Curve of O and X Waves}}
\end{figure*}

First, it is useful to review the O and X-mode waves. These waves
propagate perpendicular to the magnetic field, with cut-off and resonance
points. Here, subscripts $c$ and $p$ are used for cyclotron and
plasma frequencies, respectively, and $e$ and $i$ are used for electron
and ion species, respectively. The dispersion relation of the O-mode
wave is \cite{Miyamoto book} 
\begin{equation}
\omega^{2}=\omega_{pe}^{2}+c^{2}k_{\perp}^{2}\label{eq:O-mode Wave Dispersion Equation}
\end{equation}
where $c$ and $k_{\perp}$ respectively are light speed and wavenumber
perpendicular to the external magnetic field. It is clear from (\ref{eq:O-mode Wave Dispersion Equation})
that when $\omega=\omega_{pe}$ we have $k_{\perp}\rightarrow0$ and
propagation is cut-off, so the O-mode wave cannot propagate for $\omega<\omega_{pe}$.
The dispersion curve of the O-mode wave is shown in Figure \ref{fig:Dispersion Curve of O and X Waves-1a}
by the blue solid line. This wave also is known as an electron plasma
or Langmuir wave. The electric field component of this wave is parallel
to the external magnetic field, $\mathbf{E}\parallel\mathbf{B}_{0}$. 

The dispersion relation of the X-mode wave is \cite{Miyamoto book}
\begin{equation}
\omega^{2}=\frac{(\omega^{2}-\omega_{LH}^{2})(\omega^{2}-\omega_{UH}^{2})}{(\omega^{2}-\omega_{L}^{2})(\omega^{2}-\omega_{R}^{2})}\, c^{2}k_{\perp}^{2}\label{eq:X-mode Wave Dispersion Equation}
\end{equation}
where $\omega_{LH}$, $\omega_{UH}$, $\omega_{L}$ and $\omega_{R}$
respectively are the lower hybrid, upper hybrid, left and right-hand
polarization frequencies, and are defined as \cite{Miyamoto book}
\[
\frac{1}{\omega_{LH}^{2}}=\frac{1}{\omega_{ci}^{2}+\omega_{pi}^{2}}+\frac{1}{\mid\omega_{ci}\omega_{ce}\mid},
\]
\[
\omega_{UH}^{2}=\omega_{pe}^{2}+\omega_{ce}^{2},
\]
\[
\omega_{L}=-\frac{\omega_{ce}}{2}+\left[\left(\frac{\omega_{ce}}{2}\right)^{2}+\omega_{pe}^{2}+\mid\omega_{ci}\omega_{ce}\mid\right]^{1/2}>0,\mbox{ and}
\]
\[
\omega_{R}=\frac{\omega_{ce}}{2}+\left[\left(\frac{\omega_{ce}}{2}\right)^{2}+\omega_{pe}^{2}+\mid\omega_{ci}\omega_{ce}\mid\right]^{1/2}>0.
\]

As shown in Figure \ref{fig:Dispersion Curve of O and X Waves-1a}
by red solid lines, the X-mode wave has three branches. Based on their
phase velocities in comparison with light speed (straight dashed line),
the top and bottom branches are named respectively the fast and slow
X waves, and the middle wave for $k_{\perp}\rightarrow0$ is a fast
X wave and for $k_{\perp}\rightarrow\infty$ is a slow X wave. The
exact shape of the bottom branch of the X-mode wave is shown in \ref{fig:Dispersion Curve of O and X Waves-1b}.
It is clear from (\ref{eq:X-mode Wave Dispersion Equation}) when
$\omega=\omega_{LH}$ or $\omega=\omega_{UH}$ we have $k_{\perp}\rightarrow\infty$,
and consequently, resonance at these points. Conversely, when $\omega=\omega_{L}$
or $\omega=\omega_{R}$ we have $k_{\perp}\rightarrow0$, and consequently,
cut-off at these points. The region between $\omega_{LH}$ and $\omega_{L}$
and the region between $\omega_{UH}$ and $\omega_{R}$, shown in
Figure \ref{fig:Dispersion Curve of O and X Waves-1a}, are the cut-off
regions for which X-mode wave propagation is forbidden. The electric
field component of the X-mode wave is perpendicular to the external
magnetic field, $\mathbf{E}\perp\mathbf{B}_{0}$. 

\begin{figure}
\noindent \begin{centering}
\includegraphics[bb=0bp 0bp 600bp 715bp,scale=0.35]{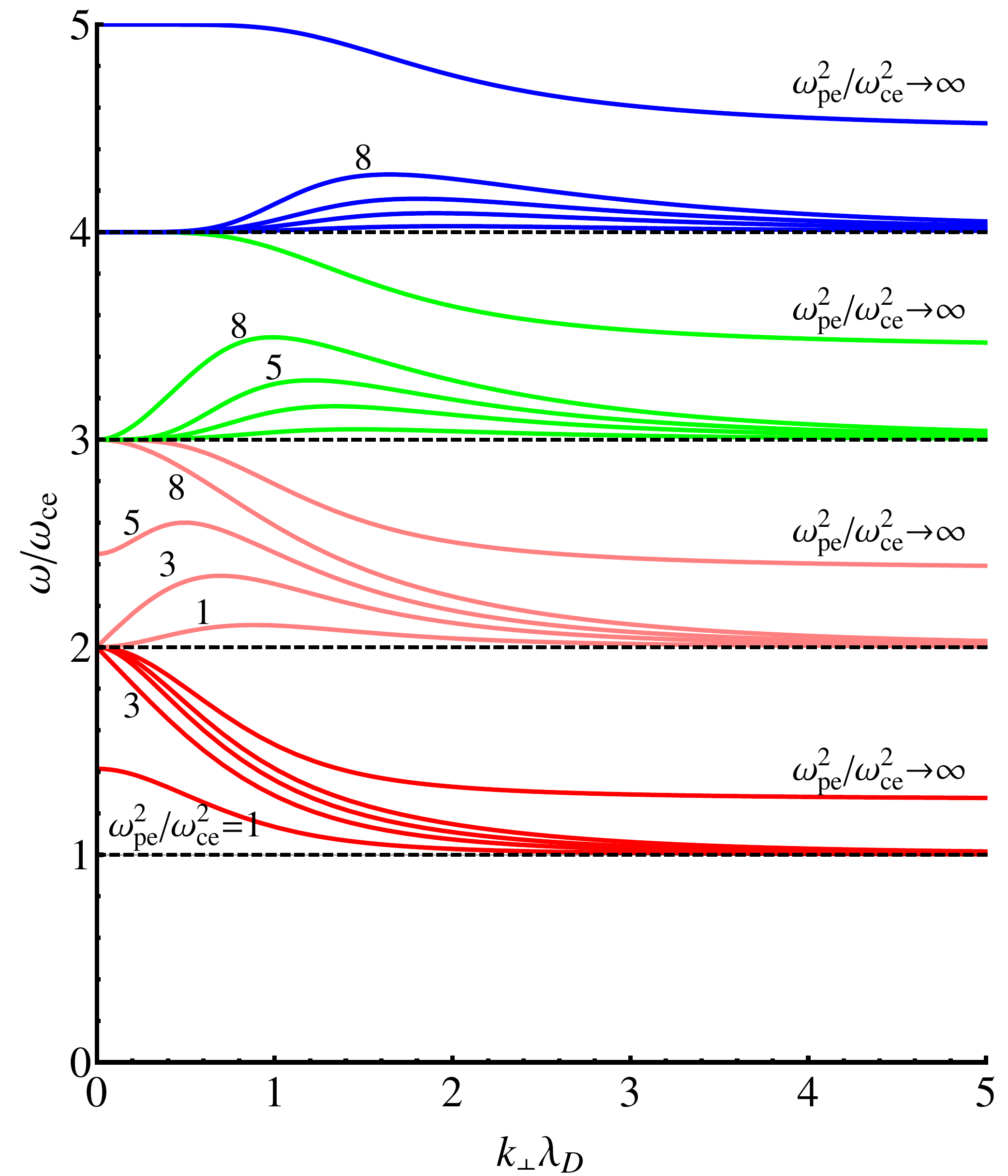}
\par\end{centering}

\caption{The dispersion curve of the EBW\label{fig:Dispersion Curve of EBW}}
\end{figure}

The dispersion of the EBW obtained for a hot plasma approximation
is \cite{Laqua Review} 
\begin{equation}
\omega^{2}=2\omega_{pe}^{2}\frac{e^{-\mu}}{\mu}\underset{n=1}{\overset{\infty}{\sum}}\frac{n^{2}I_{n}(\mu)}{1-n^{2}\omega_{ce}^{2}/\omega^{2}}\label{eq:EBW Dispersion Equation}
\end{equation}
where $\mu=\frac{1}{2}k_{\perp}^{2}r_{L,e}^{2}$ is the finite Larmor
parameter where $r_{L,e}^{2}=v_{th,e}^{2}/\omega_{ce}^{2}$ is the
electron Larmor radius with $v_{th,e}$ as the electron thermal velocity.
$I_{n}$ is the $n$th order modified Bessel function. Near the cyclotron
harmonic resonances, the dispersion relation of the EBW has an imaginary
part causing damping. So in the vicinity of the cyclotron harmonics
($\omega=n\omega_{ce}$ for integer $n$), the refractive index tends
to infinity, and as shown in Figure \ref{fig:Dispersion Curve of EBW},
the waves are strongly absorbed in these regions \cite{Crawford dispersion curve}.
On the other hand, as also shown in Figure \ref{fig:Dispersion Curve of EBW},
there is no cut-off for EBWs due to limitation of density, and even
if $\omega_{pe}/\omega_{ce}\rightarrow\infty$, the EBW would propagate
in plasma. Therefore the EBWs can penetrate into the over-dense regions
of plasma and cause local plasma heating and current drive. 

If we study plasma in a full wave model (vs WKB model), can see that
the middle branch slow X-mode wave has a longitudinal component that
becomes dominant when the wave reaches the Upper Hybrid Resonance
(UHR) point. At the UHR, the wavelength of this X-mode wave is decreasing
into the range of the Larmor radius and is leading to excitation of
the longitudinal EBW.

The EBW is generated only with collective cyclotron movements of electrons
and it is not possible to produce the EBW in vacuum space. The most
straightforward method to generate the EBW is using an electrostatic
antenna with a size around the Larmor radius, but it is clear that
using this antenna within a hot fusion plasma is not possible due
to erosion of the antenna and impurity release into the plasma, so
the practical way to excite the EBW is to use the X-mode wave with
X-B conversion.

\begin{figure}
\begin{centering}
\includegraphics[bb=0bp 0bp 768bp 769bp,scale=0.45]{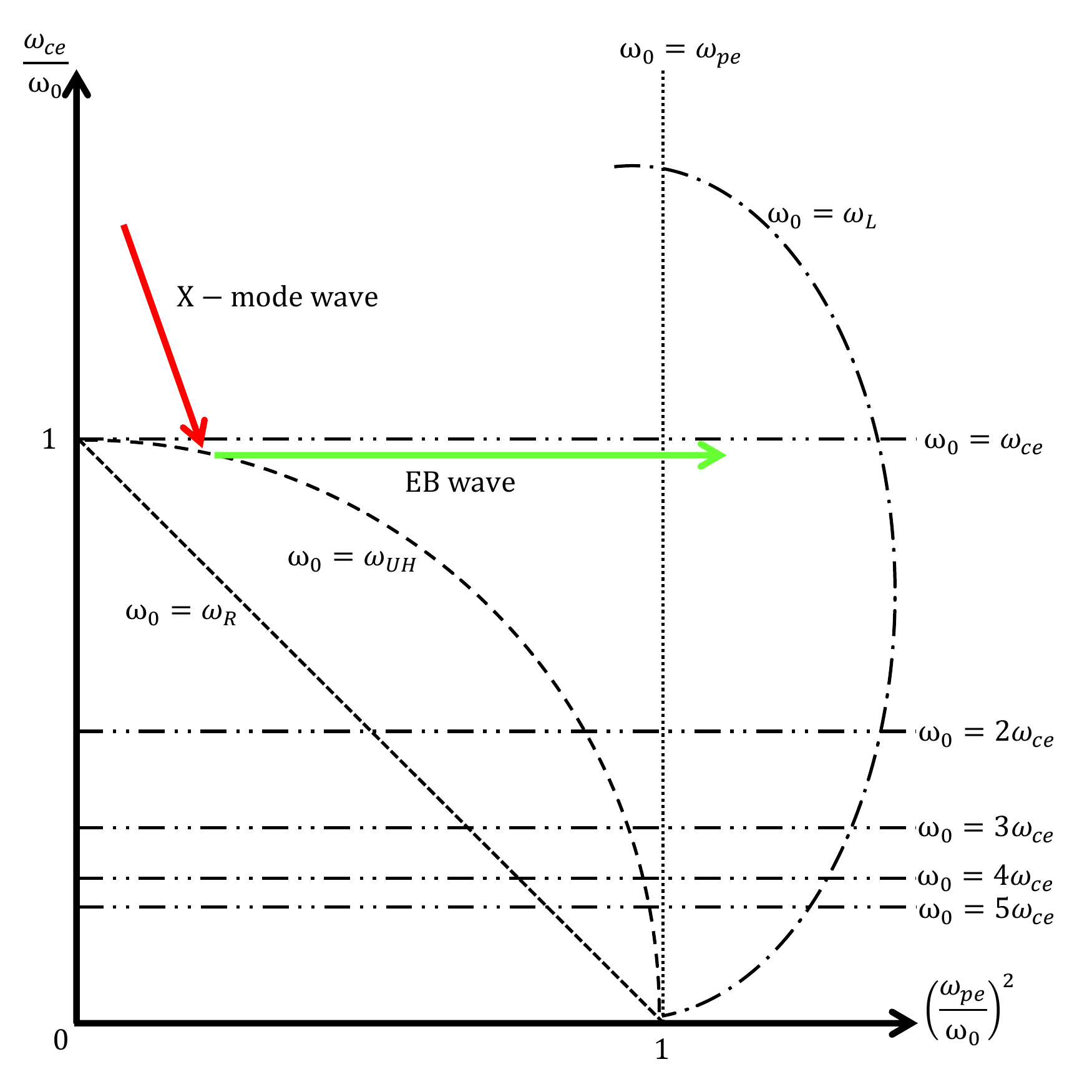}
\par\end{centering}

\caption{The schematic description of launching the slow X-mode wave from the
high magnetic field side, with X-B mode conversion at the UHR.\label{fig:CMA-HFSL}}
\end{figure}

Based on the mechanism of transmission of the extraordinary wave to
the UHR point to provide X-B conversion in spherical tokamaks, there
are three practical methods:

1. Launch the middle branch slow X-mode wave from the side with a
high magnetic field. The structure of tokamaks is such that the slow
X-mode wave would not pass through any cut-off point before reaching
the conversion point. But in STs, since the value of magnetic field
in comparison with tokamaks is very low, this condition is difficult
to meet. So this method is not popular in STs. On the other hand,
this kind of X-mode can only excite the first harmonic of the EBW,
which is in the low density region. This process is shown schematically
in Figure \ref{fig:CMA-HFSL}. An experimental work using this method
on the COMPASS-D Tokamak is \cite{HFSL}.

\begin{figure}
\noindent \begin{centering}
\includegraphics[bb=0bp 0bp 771bp 768bp,scale=0.45]{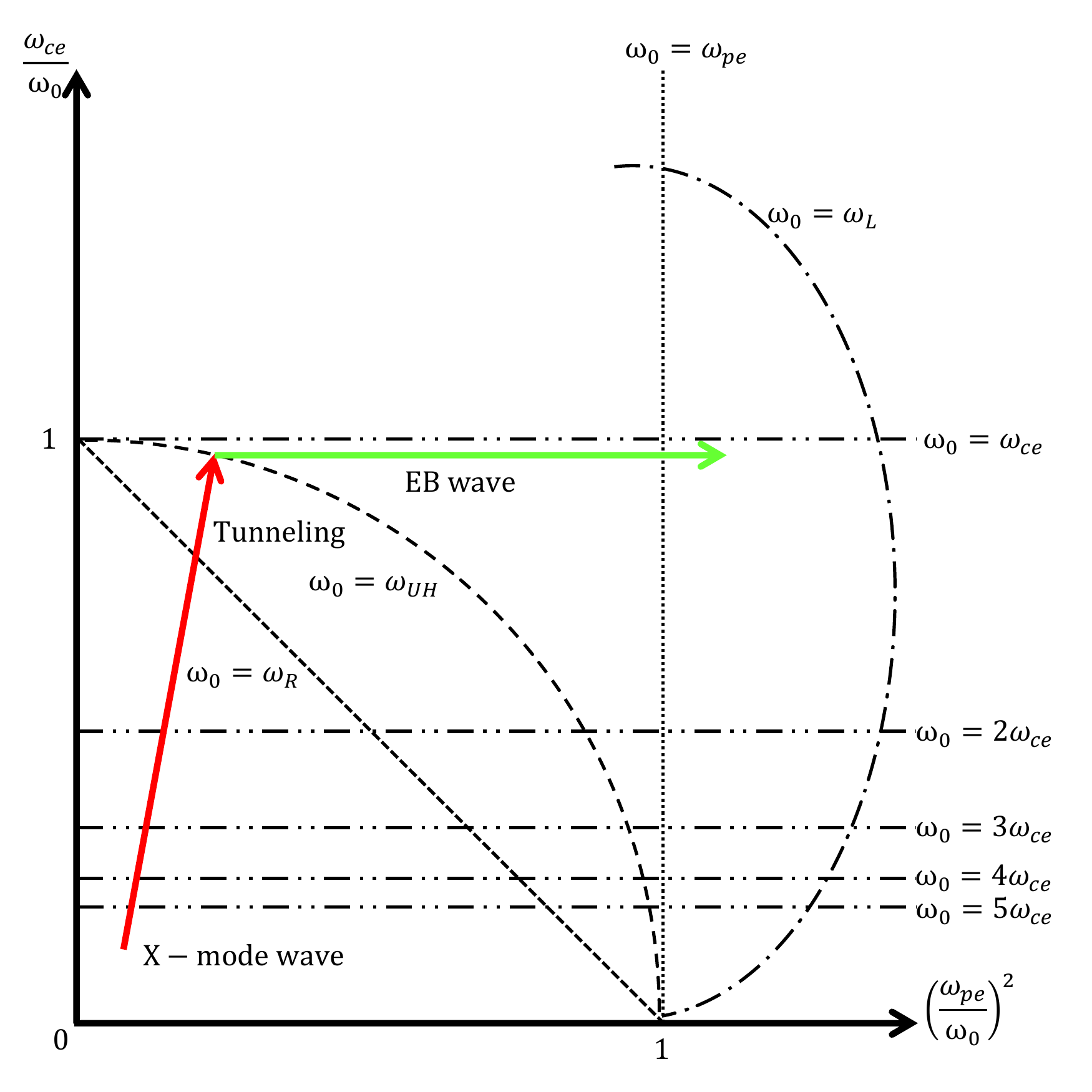}
\par\end{centering}

\caption{The schematic description of direct X-B mode conversion. In special
frequencies and angles, the fast X-mode wave is able to tunnel from
the right-hand cut-off and reach the UHR for X-B mode conversion.\label{fig:CMA-XB}}
\end{figure}

2. Launching the top branch fast X mode from the low magnetic field
side. The most important problem in this method is the existence of
a cut-off layer for this branch of the X mode wave before reaching
the UHR conversion point. It was proved theoretically and practically
that by providing good conditions in the density gradient, it is possible
for the top branch of the X-mode wave to tunnel through this layer
and convert to the slow part of the middle branch, and then this wave
will convert to the Bernstein wave. This process is shown schematically
in Figure \ref{fig:CMA-XB}. Experiments on this mode conversion scheme
have been performed at the NSTX tokamak \cite{Direct X-B}. 

\begin{figure}
\noindent \begin{centering}
\includegraphics[bb=0bp 0bp 768bp 768bp,scale=0.45]{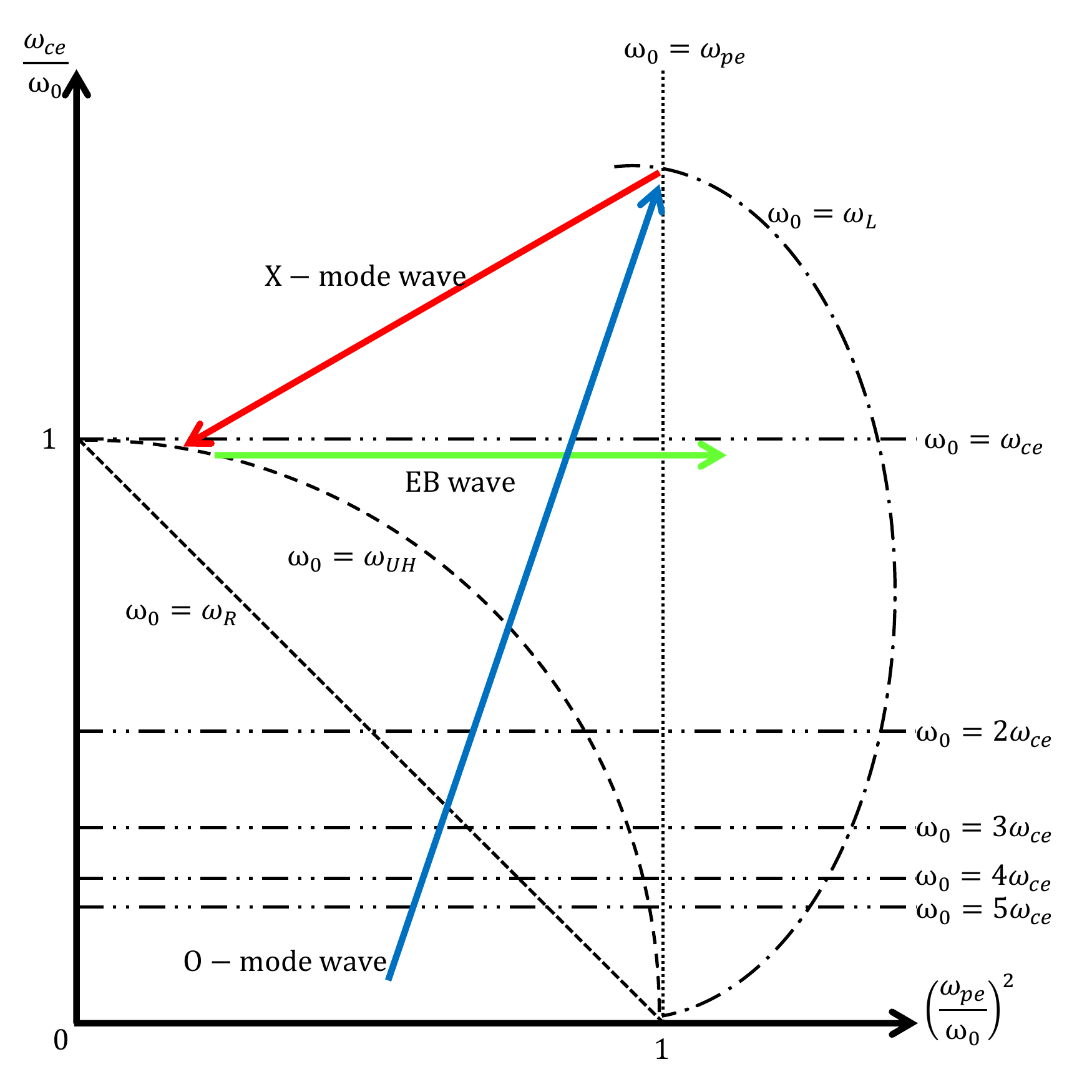}
\par\end{centering}

\caption{The schematic description of O-X-B double mode conversion. The O-mode
wave reaches the cut-off point when $\omega=\omega_{pe}$, and the
O-X conversion occurs under this condition for incidence angle and
gradient of density: $\theta=\theta_{opt}$ and $k_{0}L_{n}\geqslant10$.\label{fig:CMA-OXB}}
\end{figure}

3. Launching an O-mode wave from vacuum to produce an internal slow
X-mode wave taking part in X-B conversion. This is a double O-X-B
mode conversion process. The incoming O-mode wave, after reaching
the cut-off point at $\omega=\omega_{pe}$, is converted to a slow
X-mode wave propagating toward the UHR under special conditions. This
special condition is provided if the incoming O-mode wave is cut-off
at the point of the cut-off of the slow X-mode wave ($\omega=\omega_{L}$).
This situation leads to coincidence of O and X waves at the same point.
The conversion to the EBW mode proceeds as before, but now the higher
harmonics of the cyclotron frequency are reachable. This process is
shown schematically in Figure \ref{fig:CMA-OXB}. The efficiency of
the first O-X mode conversion is strongly depend upon the direction
of propagation with respect to the external magnetic field \cite{Optimum Angle Preinhaelter,Optimum Angle Hansen}.
The optimized angle between the wave-vector and the external magnetic
field is 
\begin{equation}
\theta_{opt}=\arccos\sqrt{\frac{\omega_{ce0}}{\omega_{ce0}+\omega_{pe0}}}\label{eq:Optimum Angle}
\end{equation}
where $\omega_{ce0}$ and $\omega_{pe0}$ respectively are the cyclotron
and plasma frequencies at the cut-off point of the O-mode wave. For
this optimal angle, both the X and O waves are coincident at the cut-off
of the O-mode wave, and have the same phase and group velocity. In
this case, the power is transferred between them without any reflection.
On the other hand, the transmission of energy also depends on the
gradient of density, shown by the inhomogeneity scale length of density
$L_{n}$. It has been shown that the best condition to have a reasonable
transmission is $k_{0}L_{n}\geqslant10$, where $k_{0}$ is the wave-number
of the O-mode wave in vacuum \cite{Scale Length}.

\subsection{Particle-In-Cell Model}

Plasma simulation can be used to study the interaction between plasma
particles and electromagnetic fields. For this purpose, two different
systems are appropriate: the Vlasov\textendash{}Maxwell system and
the Lorentz\textendash{}Maxwell system \cite{Birdsall Book}. In Vlasov\textendash{}Maxwell
solvers, the evolution of moments of the particle distribution function
is considered on a grid in phase space, and finally the required quantities
are obtained via the coupling between Maxwell's equations and the
currents and charge densities given by the temporally and spatially
varying distribution function. In the second method, the movement
of super particles due to the Lorentz force is considered on a grid
in space, and finally the required quantities are obtained via the
coupling between Maxwell's equations and the currents and charge densities
obtained from particle positions and velocities. One example of a
Lotentz-Maxwell system is the particle-in-cell (PIC) method.

The PIC model, due to the use of fundamental equations with only statistical
approximation, often is useful for considering the nonlinear effects
and the plasma collective behavior, which can be included self-consistently
by coupling charged particles to the Maxwell equations via the source
terms. Moreover, the study of relativistic effects via the relativistic
Lorentz equation and collisional effects via Monte Carlo collisions
is possible in the PIC method.

\begin{figure*}
\noindent \begin{centering}
\includegraphics[scale=0.5]{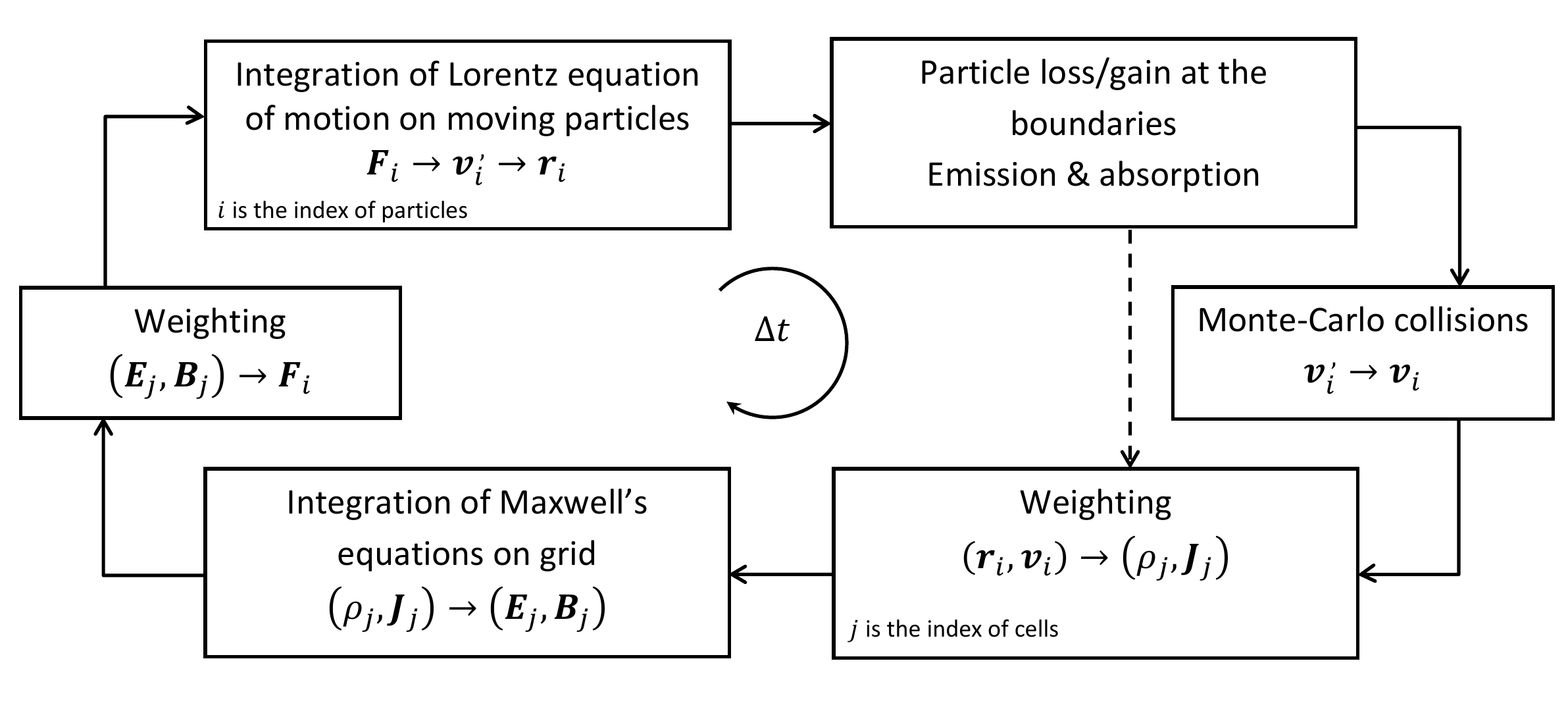}
\par\end{centering}

\caption{Calculating the positions and velocities of particles iterated in
each time step\label{fig:PIC Flowchart}}
\end{figure*}

The process cycle of calculating of positions and velocities of particles
each time step is shown schematically in Figure \ref{fig:PIC Flowchart}.
First, the velocities of the particles are obtained by integration
of the Lorentz equations of motion, and then the positions are obtained
by integration of the respective velocities. Next, the particle losses
due to absorption and gains due to emission are considered at the
boundaries. In collisional models, the perturbation of velocities
by elastic and inelastic collisions is considered. Then, the charge
densities and currents are obtained from the positions and velocities
of particles. These densities and currents are used to compute electromagnetic
fields on the spatial grid by the discretized Maxwell's equations.
Finally, the Lorentz force due to these fields is used to obtain the
positions and velocities of the particles in the next step time.

\section{Details of Modeling}

\subsection{Physical System}

\begin{figure}
\noindent \begin{centering}
\includegraphics[scale=0.4]{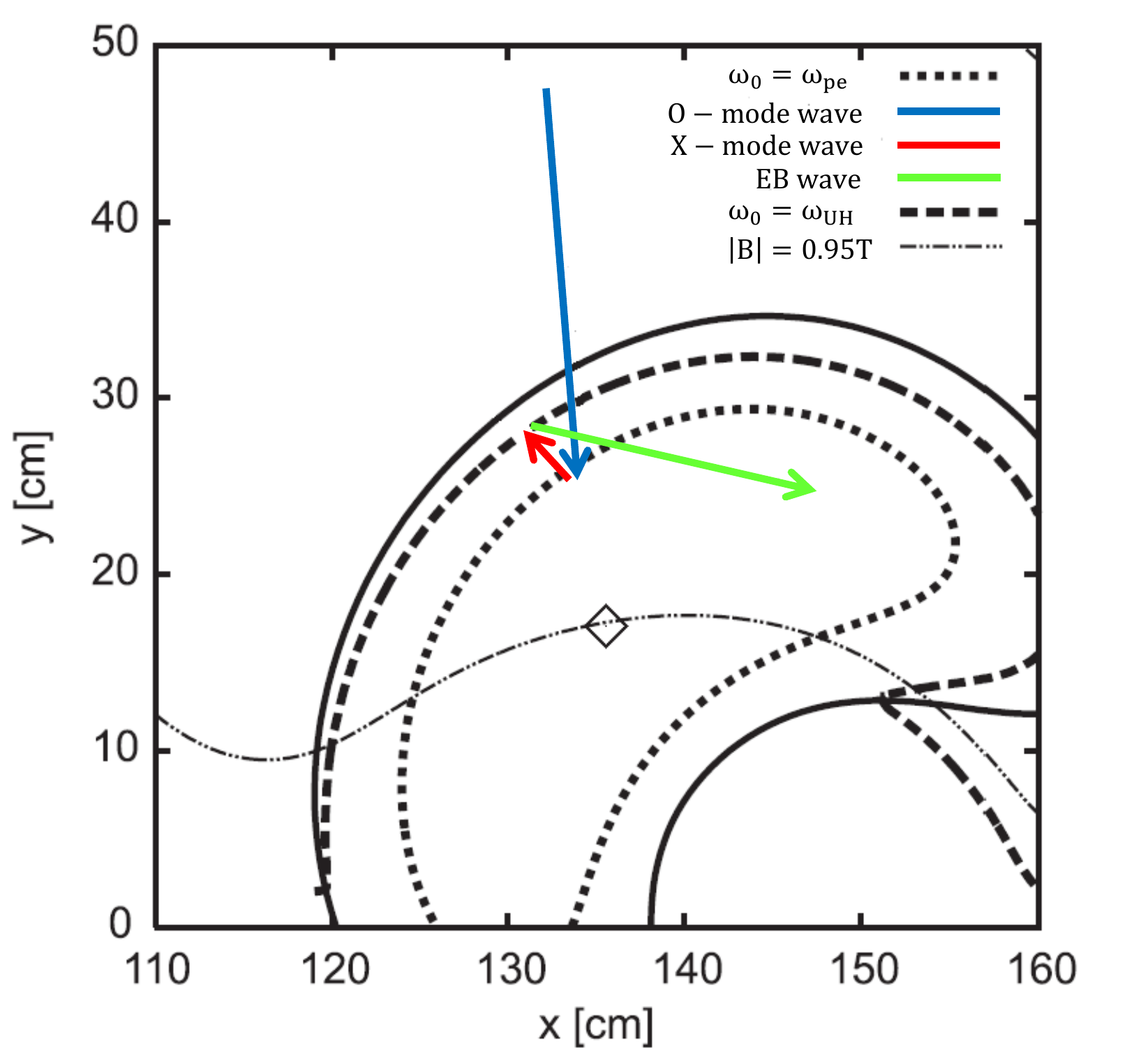}
\par\end{centering}

\caption{Poloidal cross section of TJ-II \cite{TJ-II kohn simulation}. \label{fig:Poloidal Cross Section of TJ-II}}
\end{figure}

The most suitable device for our computational simulation is a stellarator
operating in Madrid, Spain called TJ-II \cite{TJ-II-1,TJ-II-2,TJ-II-3}.
This stellarator is a medium-sized flexible heliac with the following
parameters: major radius $R{}_{0}=1.5\,\textrm{m}$, minor radius
$a=0.2\,\textrm{m}$, and magnetic field strength on axis around $B_{0}=0.95\,\textrm{T}$.
The plasma is generated and heated by launching electron cyclotron
microwaves at the second harmonic, about $f=53.2\,\textrm{GHz}$.
These microwaves are provided by two $300\,\textrm{kW}$ gyrotrons.
Electron cyclotron resonance heating is not possible for dense regions
more than $n_{e}\approx1.7\times10^{19}\,\textnormal{m}^{-3}$ because
the second harmonic is limited by the cut-off density for this frequency.
On the other hand, by providing two $800\,\textrm{kW}$ neutral beam
injection systems, it is possible to heat plasma above this limit
again with electron cyclotron resonance heating, but now via electron
Bernstein waves \cite{TJ-II EBW}. The applied scenario here is via
O-X-B double mode conversion. The O-mode wave is launched from the
low field side at the first harmonic $f_{0}=28\,\text{GHz}$. As shown
schematically in Figure \ref{fig:Poloidal Cross Section of TJ-II},
the O-mode wave cannot propagate through the cutoff layer, and so
is converted to the X-mode wave upon reflection. The efficiency of
conversion depends on the angle of the O-mode wave with respect to
the external magnetic field, and on the density gradient scale length.
The optimum angle is determined via \ref{eq:Optimum Angle} and the
experimental density gradient scale length is $k_{0}L_{n}\approx30$
\cite{TJ-II kohn simulation}. The converted X-mode wave propagates
backward, and upon reaching the UHR layer, is converted to a quasi
electrostatic Bernstein wave. This wave can penetrate deeply into
high density region in the stellarator core plasma. 

\subsection{Simulation of Physical Model}

XOOPIC is a particle simulation code which has been used for modeling
dense magnetized plasma. The XOOPIC (X11-based object-oriented particle-in-cell)
code is an advanced two dimensional PIC code with high flexibility
and extensibility due to the object oriented paradigm (OOP) \cite{XOOPIC OOP}.
In XOOPIC, we can simulate our system in slab or axisymmetric cylindrical
geometries. The velocities and electromagnetic field vectors are considered
in three dimensions, with spatial variation in two dimensions. XOOPIC
includes the XGRAFIX interface that allows the user to display multiple
diagnostics and view them as they evolve in time \cite{XOOPIC XGRAFIX}.
For obtaining the electromagnetic field, Maxwell's equation are solved
by enforcing conservation laws in discrete finite volumes\cite{XOOPIC field solvers}.
The density and current sources of electromagnetic field are provided
by solving the relativistic equations of motion. These relativistic
equations have been descretized by the relativistic time-centered
Boris method \cite{PIC Review}. For eliminating the problem of large
angular displacements and singularity of angular momentum near the
origin in axisymmetric coordinates, the position of particles is updated
in a rotated Cartesian frame \cite{Birdsall Book}. The incident wave
is generated by a wave source implementing a surface impedance boundary
condition (SIBC) \cite{XOOPIC ExitPort}.

\begin{figure*}
\noindent \begin{centering}
\includegraphics[scale=0.8]{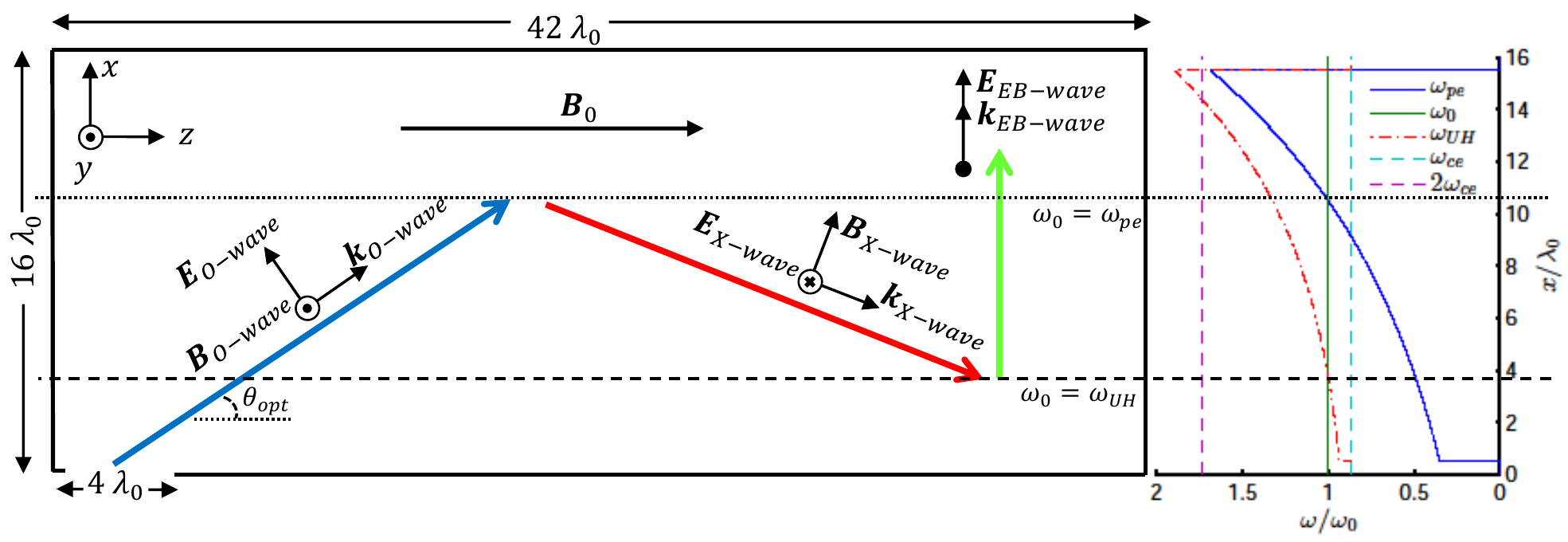}
\par\end{centering}

\caption{Schematic description of wave propagation into the plasma in the planar
simulation coordinate system. Diagram of spatial dependence of plasma,
cyclotron and upper hybrid frequencies in comparison with frequency
of the incident wave is plotted on the right.\label{fig:System simulation}}
\end{figure*}

As has been shown schematically in Figure \ref{fig:System simulation},
we have modeled the wave conversion region using planar geometry.
Its width and length respectively are $L_{x}=16\,\lambda_{0}$ and
$L_{z}=42\,\lambda_{0}$ where $\lambda_{0}$ is the vacuum wavelength
of the incident wave. For $f_{0}=28\,\textrm{GHz}$, $L_{x}=17\,\textrm{cm}$
and $L_{z}=45\,\textrm{cm}$. The number of cells in x and y directions
are $n_{x}=4096$ and $n_{z}=1344$, respectively. Then the size of
the cells is about $42\,\mu\text{m}\times335\,\mu\text{m}$, for an
aspect ratio of about $8$. The length of the input port for the incident
wave is $4\,\lambda_{0}$, and comprising $128$ cells. The propagation
direction of the generated electromagnetic waves for this input port
in XOOPIC is perpendicular to the emitter walls \cite{XOOPIC ExitPort}.
In order to generate an oblique emitted wave, we divided the port
into $128$ sub-ports. By adjusting the phase and amplitude of each
sub-port, ultimately we established an oblique Gaussian wave propagating
in the desired angle and spatial amplitude dependence. For this simulation
based on \ref{eq:Optimum Angle}, we have $\theta_{opt}=47^{\circ}$.

Since with respect to the width of the simulation region, the magnetic
field variation is very small, we assumed that magnetic field is constant
and according to previous simulations \cite{TJ-II kohn simulation}
we used $B_{0}=0.87\,\textrm{T}$. The functional behavior of density
is exponential so $n(x)$ is determined as
\begin{equation}
n(x)=n_{0}e^{\frac{x-x_{O-cutoff}}{L_{n}}}\label{eq:density}
\end{equation}
where $n_{0}$ is the density at O-mode wave cut-off position, $x_{O-cutoff}$,
and $L_{n}=\left[\frac{dln\left[n(x)\right]}{dx}\right]^{-1}$ is
the density inhomogeneity scale. The value of $L_{n}$, obtained from
$k_{0}L_{n}=30$ \cite{TJ-II kohn simulation}, is about $5.11\,\textrm{cm}$,
where $k_{0}$ is the wavenumber of incident wave. The locations of
the upper hybrid resonance layer and O cut-off layer, shown in Figure
\ref{fig:System simulation}, are given by the intersection points
of the upper hybrid and plasma frequencies with the constant frequency
of incident wave. The values of these quantities are $x_{UHR}=3.747\,\lambda_{0}$
and $x_{O-cutoff}=10.5\,\lambda_{0}$ respectively. The diagram of
the local frequencies obtained from the magnetic field and density
functions and the resonance and cut off layers are shown on the right
in Figure \ref{fig:System simulation}.

\section{Results}

\begin{figure}
\begin{centering}
\includegraphics[scale=0.4]{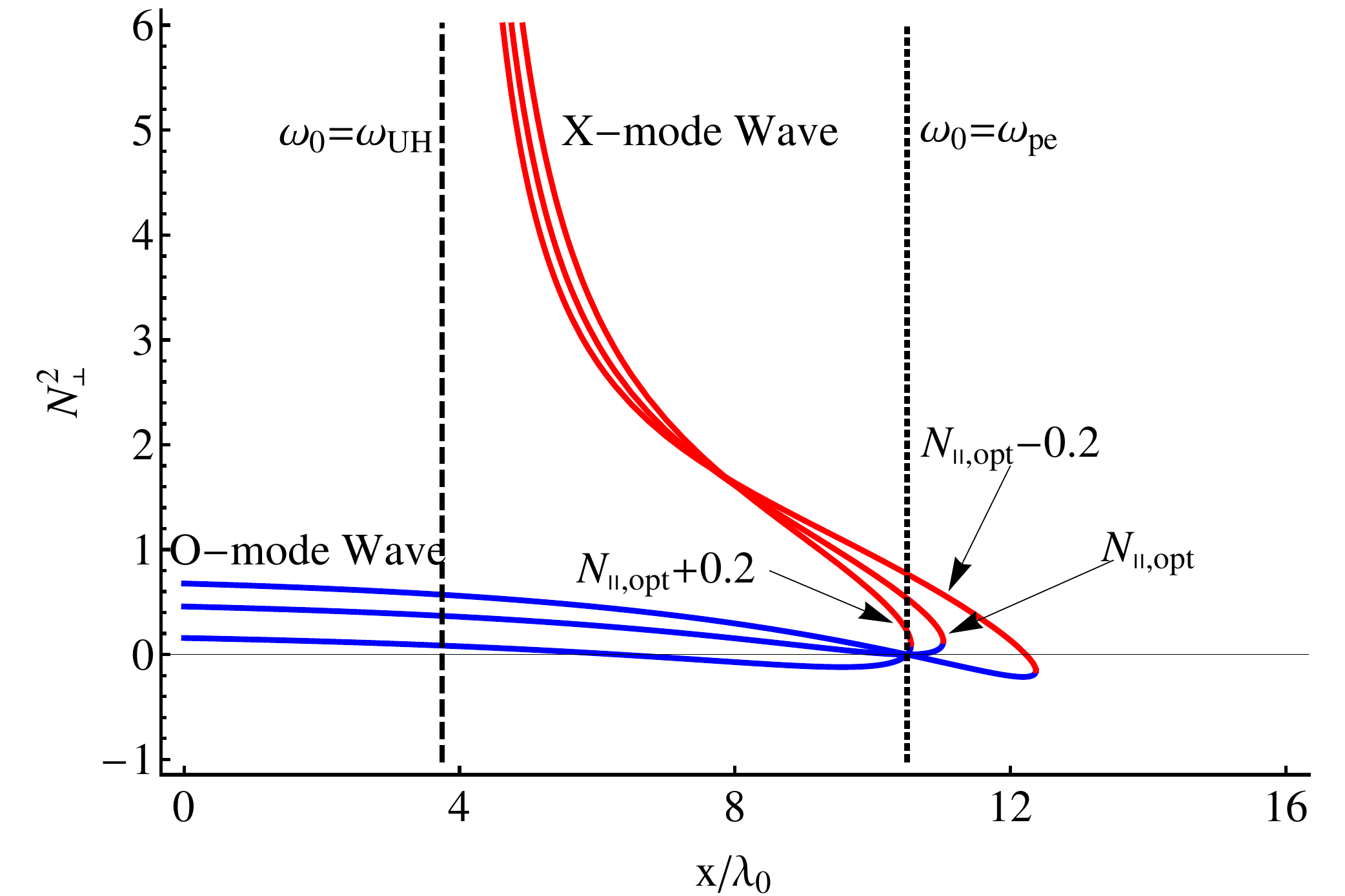}
\par\end{centering}

\caption{Perpendicular component of refractive index for a launched O-mode
wave and a reflected X-mode wave in the x direction of the simulation
region for TJ-II conditions for different launch angles. The O-mode
coincides with the X-mode near the cutoff layer at $\omega_{0}=\omega_{pe}$.
The blue and red lines are O-mode and X-mode waves, respectively.\label{fig:optimum parallel N}}
\end{figure}

To verify the optimum angle $\theta_{opt}=47^{\circ}$, the perpendicular
refractive index of O and X-mode waves along the x direction is obtained
for varying parallel refractive index $N_{\parallel}=\cos\theta$.
For this purpose, the wave equation\textbf{\textit{
\begin{equation}
N\times(N\times E)+KE=0\label{eq:wave equation}
\end{equation}
}}is used where $N$ and $K$ are the total refractive index and the
'cold' dielectric tensor, respectively \cite{Miyamoto book,Laqua Review}.
As shown in Figure \ref{fig:optimum parallel N}, by using the conditions
of TJ-II, for other values of the parallel component of refractive
index larger and smaller than $N_{\parallel,opt}$, the perpendicular
component of refractive index $N_{\perp}$ of O or X-mode waves has
an imaginary part that reduce the chance of conversion. But for $N_{\parallel,opt}=\cos\theta_{opt}=0.68$,
both waves have positive $N_{\perp}^{2}$ at the conversion point,
and because they coincidence at the cutoff layer, the conversion is
the most probable. Thus $\theta_{opt}=47^{\circ}$ can be confirmed
by PIC simulation.

\begin{figure*}
\subfloat[\label{fig:OXB-TJ-II-a}]{\centering{}\includegraphics[scale=0.35]{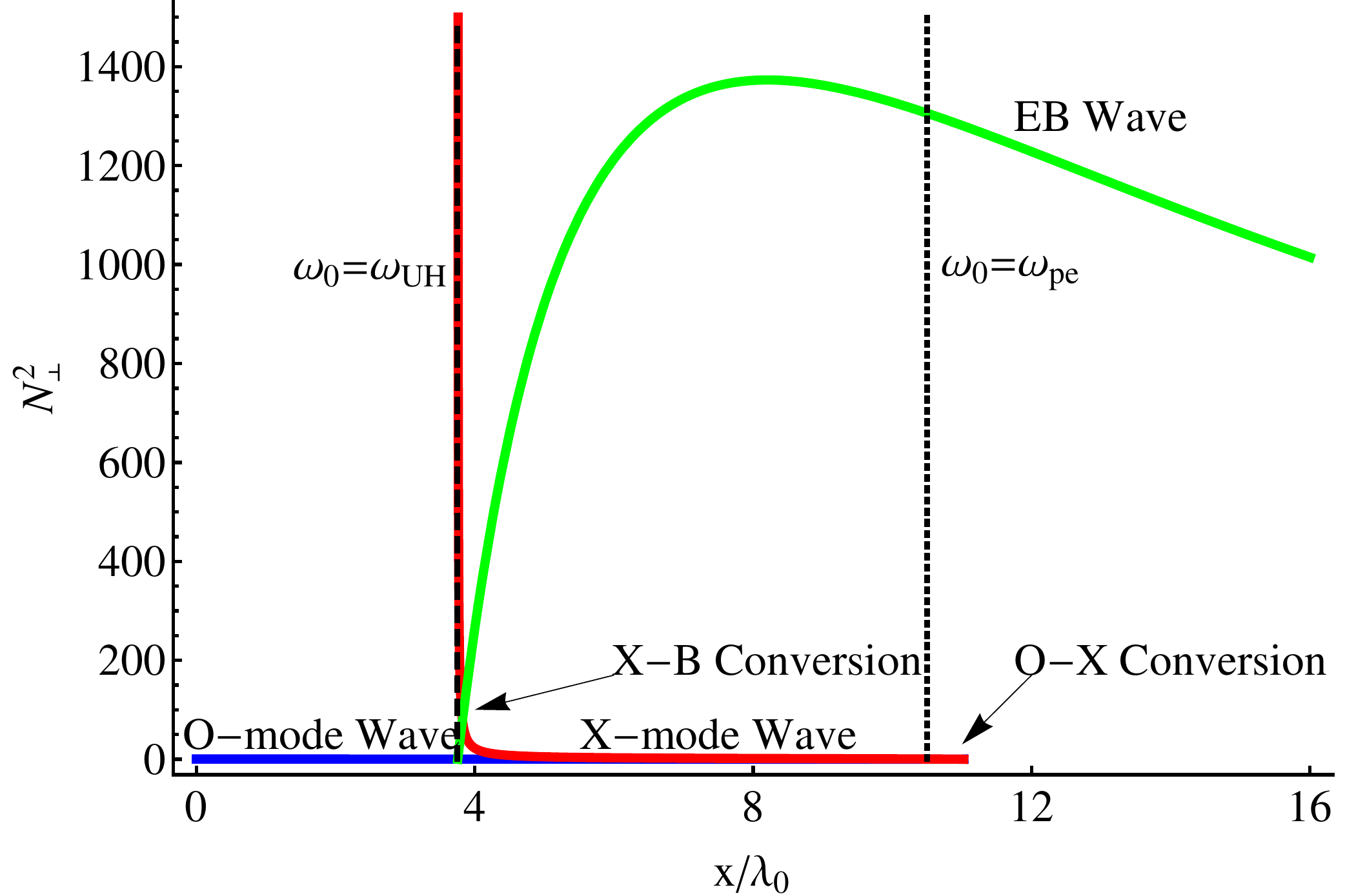}}\subfloat[\label{fig:OXB-TJ-II-b}]{\centering{}\includegraphics[scale=0.35]{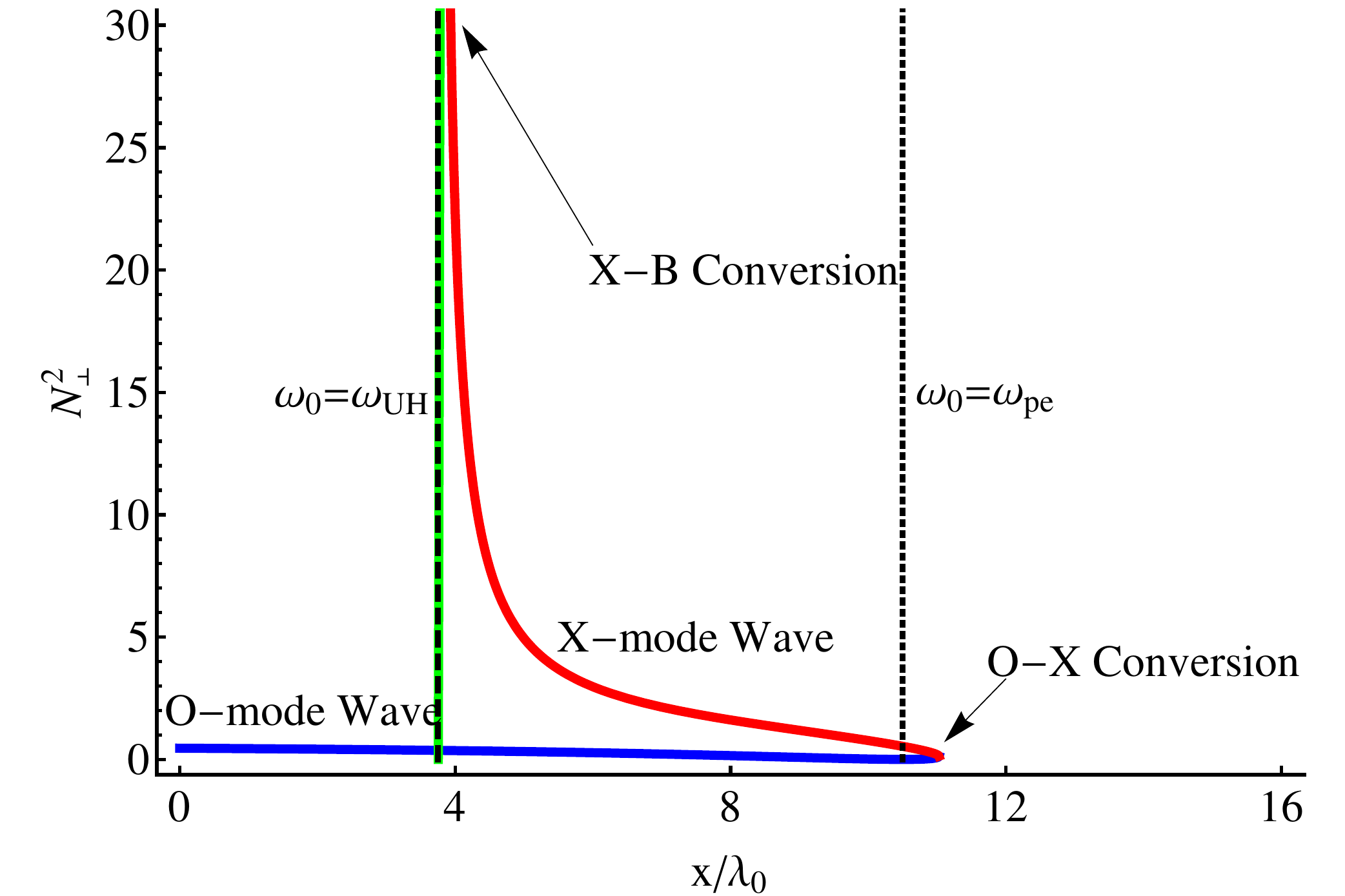}}\caption{The mechanism of O-X-B double conversion by considering the perpendicular
component of refractive index for the launched O-mode wave (blue line),
reflected X-mode wave (red line) and EBW (green line) in the x direction
of the simulation region for TJ-II conditions in (a) full and (b)
short range. The O-X conversion occurs near the cutoff layer at $\omega_{0}=\omega_{pe}$,
and X-B conversion occurs at the UHR layer at $\omega_{0}=\omega_{UH}$.
\label{fig:OXB-TJ-II}}
\end{figure*}

On the other hand, it is very important to verify penetration and
propagation of the EBW in the x direction and beyond the cutoff layer.
For this purpose, we can use Equation \ref{eq:wave equation}, but
now with the 'hot' dielectric tensor \cite{Laqua Review} to obtain
the refractive index of the EBW. As shown in Figure \ref{fig:OXB-TJ-II-a},
for TJ-II parameters, the EBW can propagate to dense regions after
passing the cutoff layer of the O-mode wave. This corresponds to the
lack of density limitation shown in Figure \ref{fig:Dispersion Curve of EBW}.
As detailed in Figure \ref{fig:OXB-TJ-II-b}, the O-X conversion happens
near the O cutoff layer, and the X-B conversion occurs exactly in
the UHR layer when the reflected X-mode wave reaches this layer. When
$N_{\perp}$ of the X-mode wave and the EBW are equal, the X-mode
wave is converted to the EBW.

The PIC simulation was performed on a cluster of parallel computers
using the MPI mechanism. The results of this simulation are divided
into the following three sections:

\subsection{Generation of the O-mode wave}

\begin{figure*}[tp]
\subfloat[$E_{x}$]{\includegraphics[scale=0.5]{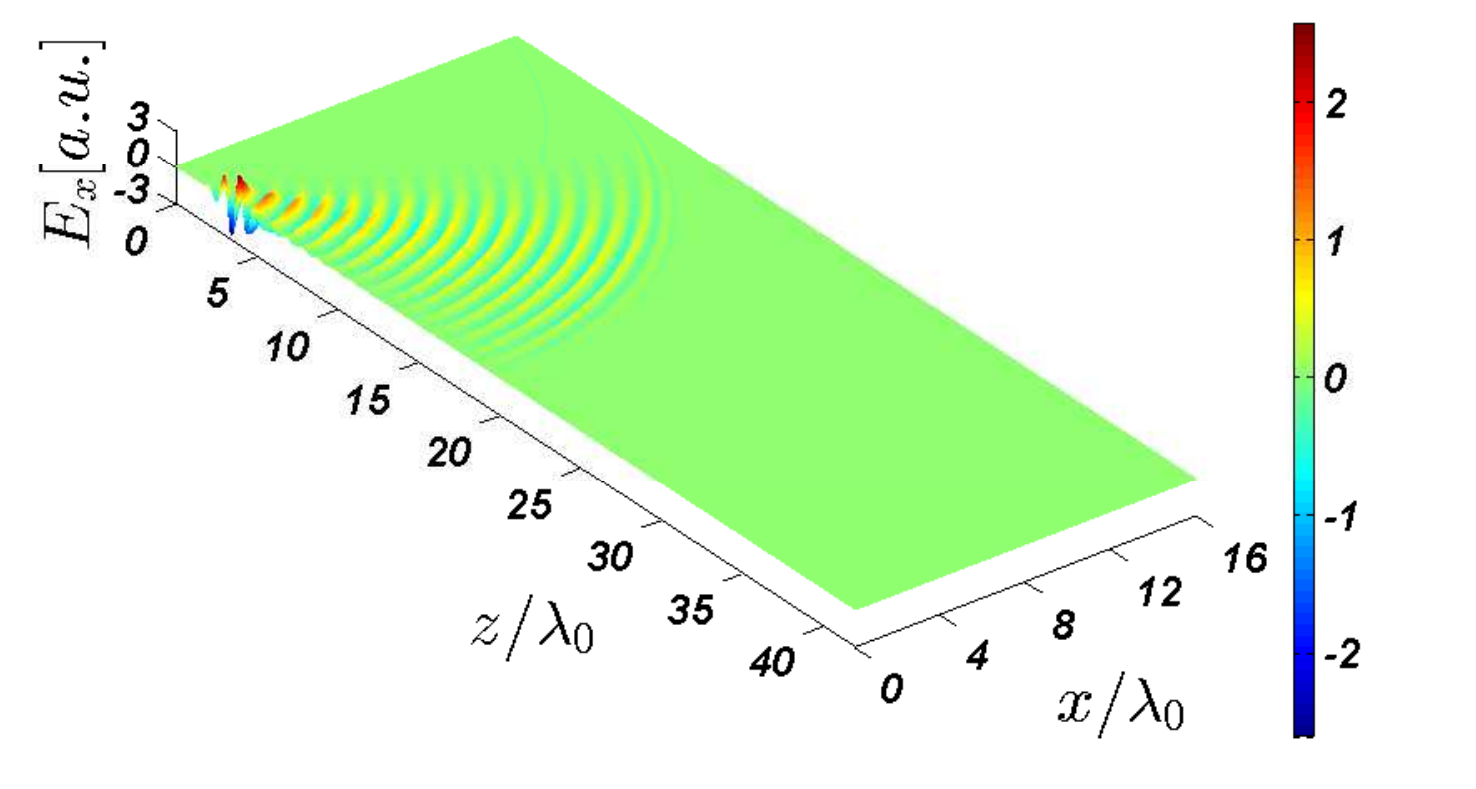}}\subfloat[$B_{x}$]{\includegraphics[scale=0.5]{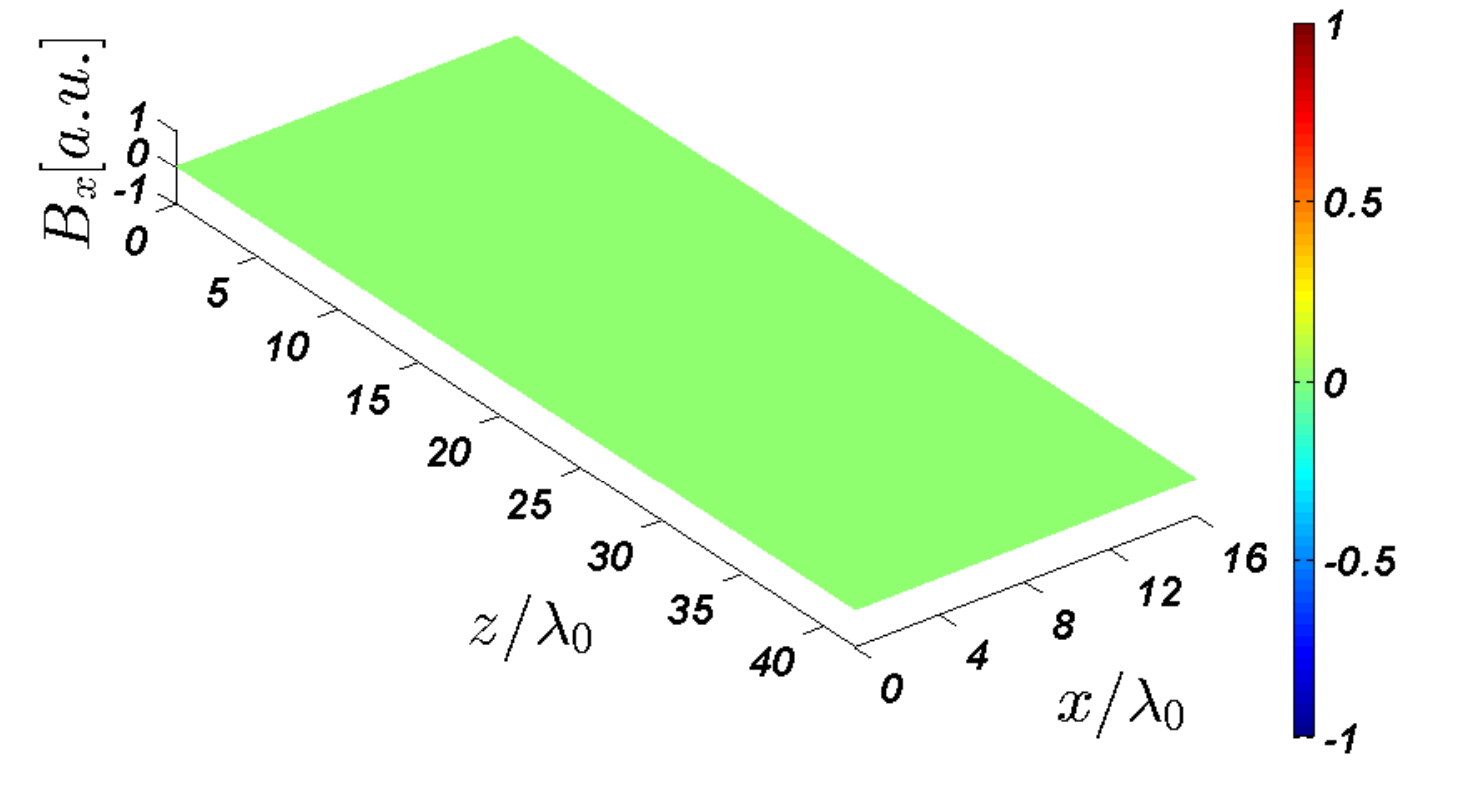}}\medskip{}
\subfloat[$E_{y}$]{\includegraphics[scale=0.5]{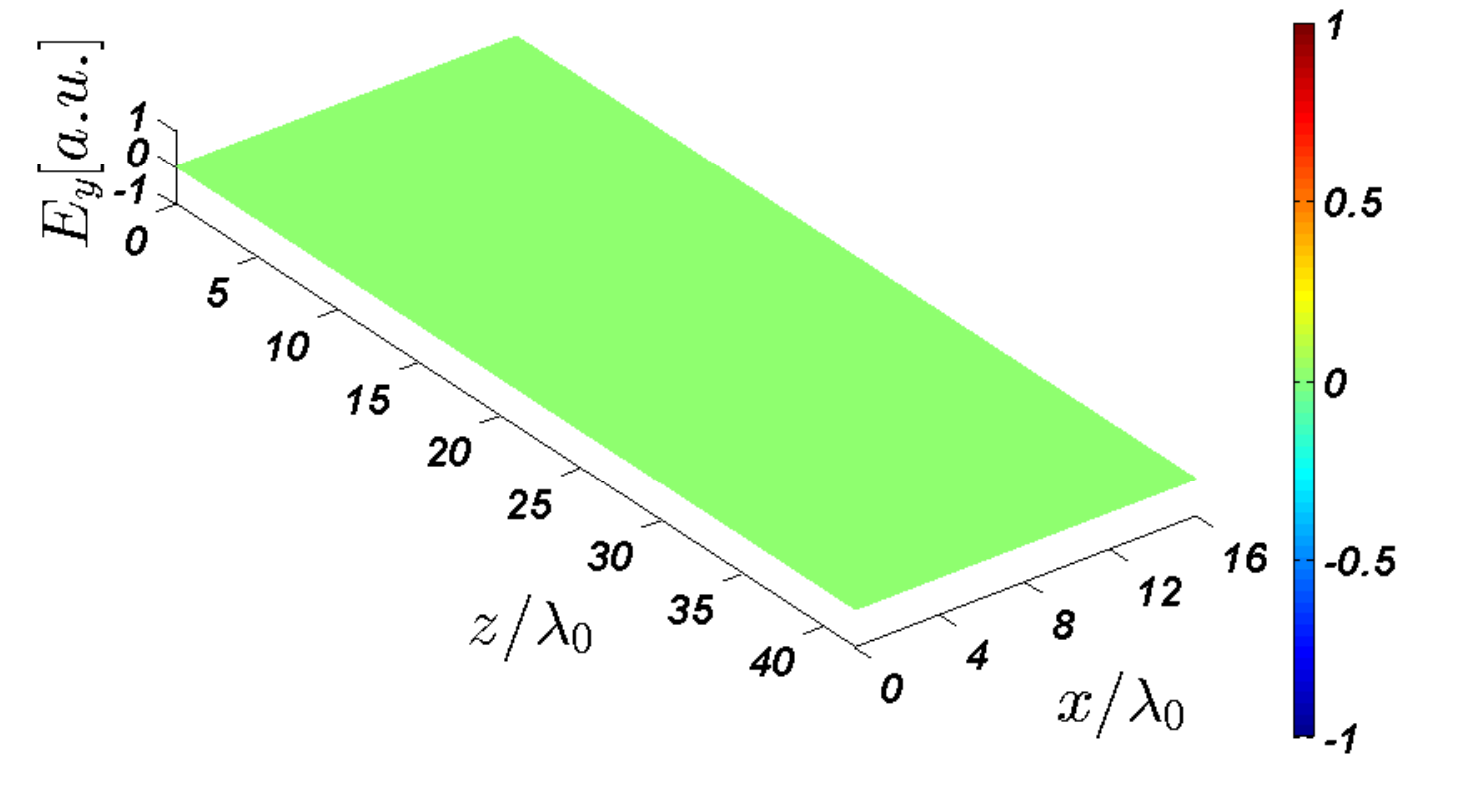}}\subfloat[$B_{y}$]{\includegraphics[scale=0.5]{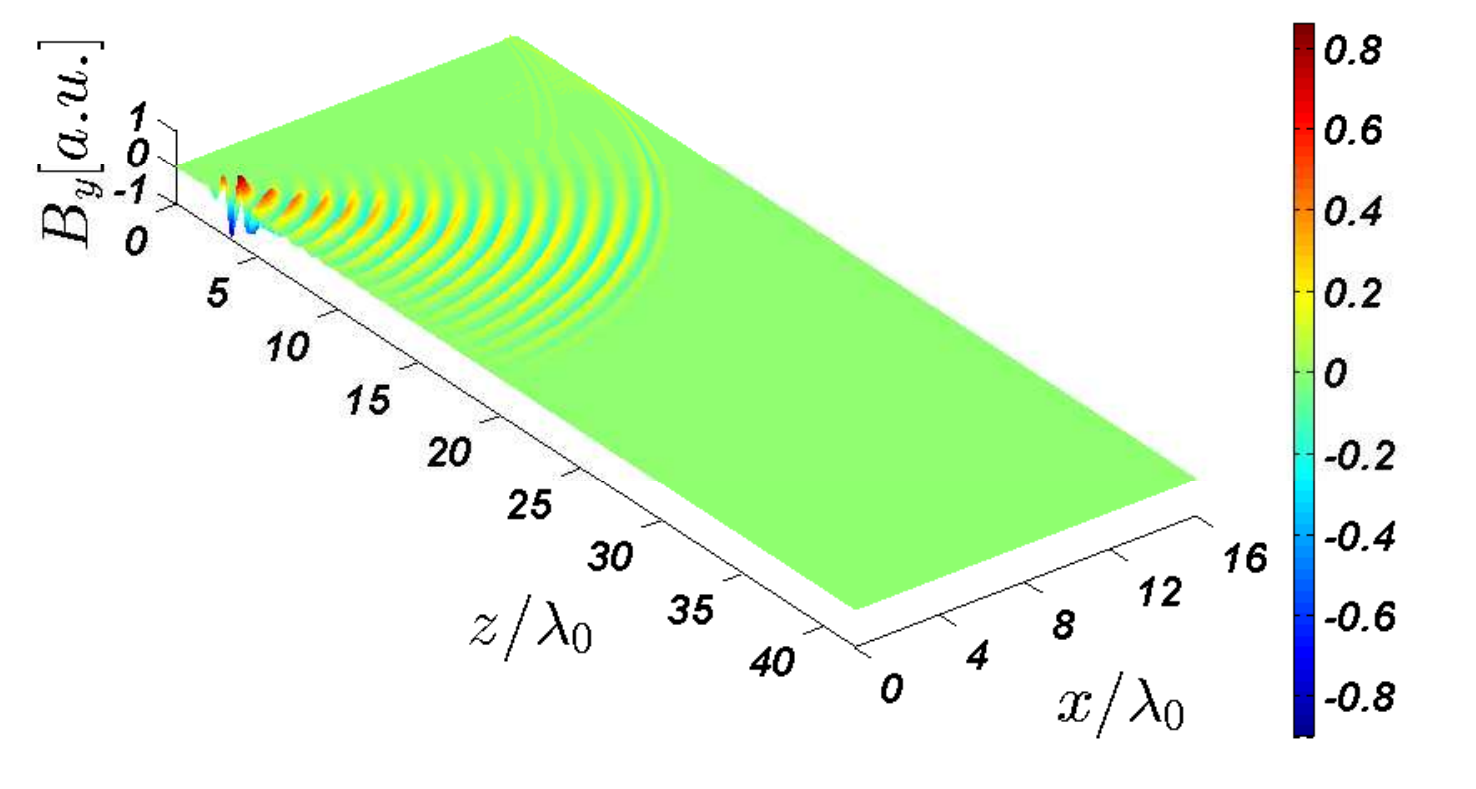}}\medskip{}
\subfloat[$E_{z}$]{\includegraphics[scale=0.5]{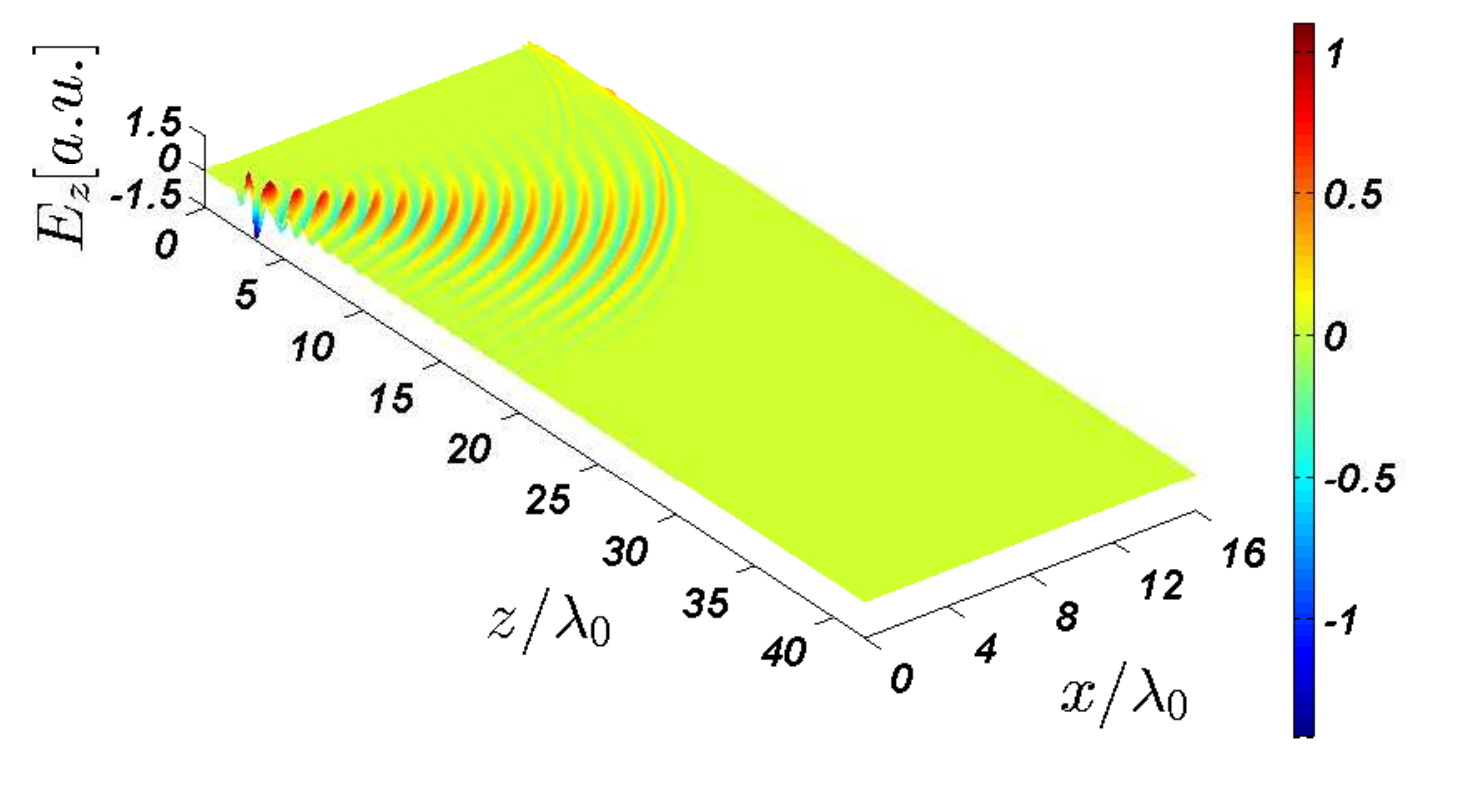}}\subfloat[$B_{z}$]{\includegraphics[scale=0.5]{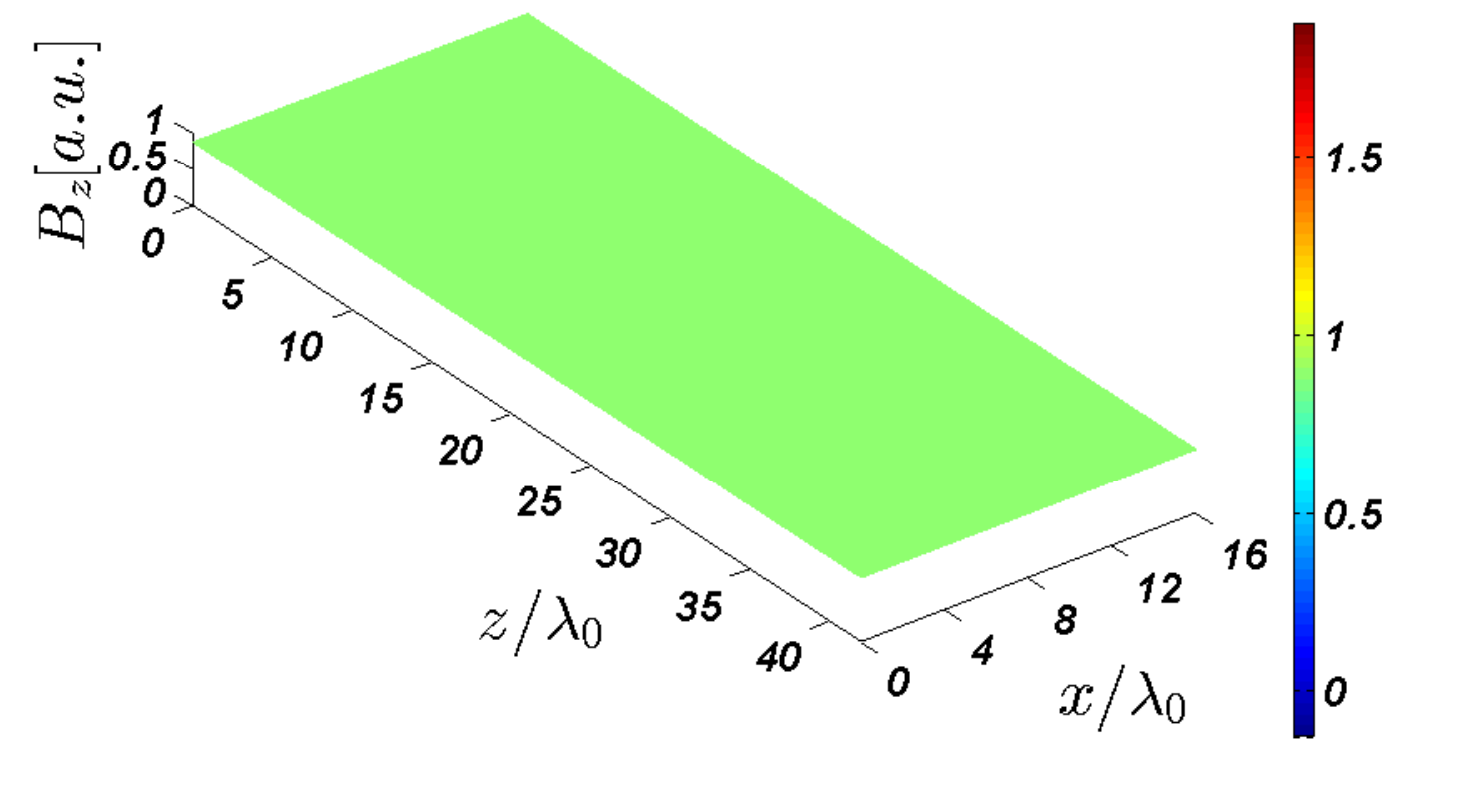}}\caption{Distribution of electric and magnetic field components of the launched
O-mode wave within the simulated region without plasma particles at
time $t=16\, T_{wave}$, where $T_{wave}=1/f_{0}$ is time period
of the launched wave.\label{fig:O wave components}}
\end{figure*}

According to Figure \ref{fig:System simulation} and the value of
the launch angle, we can see that the incident O-mode wave should
have the $E_{x}$ and $E_{z}$ components of electric field and the
$B_{y}$ component of magnetic field. As shown in Figure \ref{fig:O wave components},
we have adjusted the launched wave manually to be an O-mode wave.
The incident wave just has $E_{x}$, $E_{z}$ and $B_{y}$ components
and other components are zero, indicating the presence of the O-mode
wave.

\subsection{Observation of the reflected X-mode wave}

\begin{figure*}
\subfloat[$\left|\vec{E}\right|$]{\includegraphics[scale=0.5]{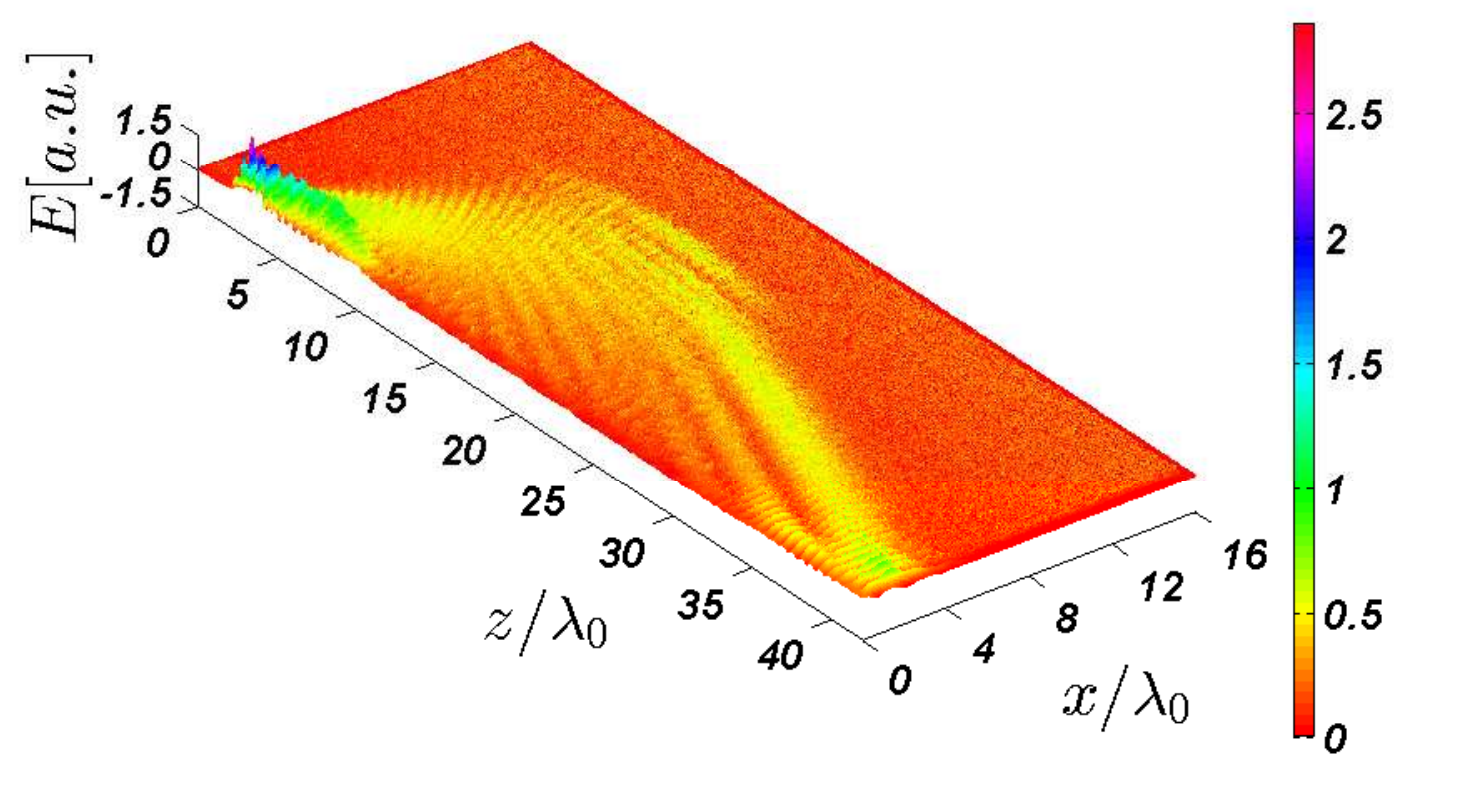}}\subfloat[$\left|\vec{B}\right|$]{\includegraphics[scale=0.5]{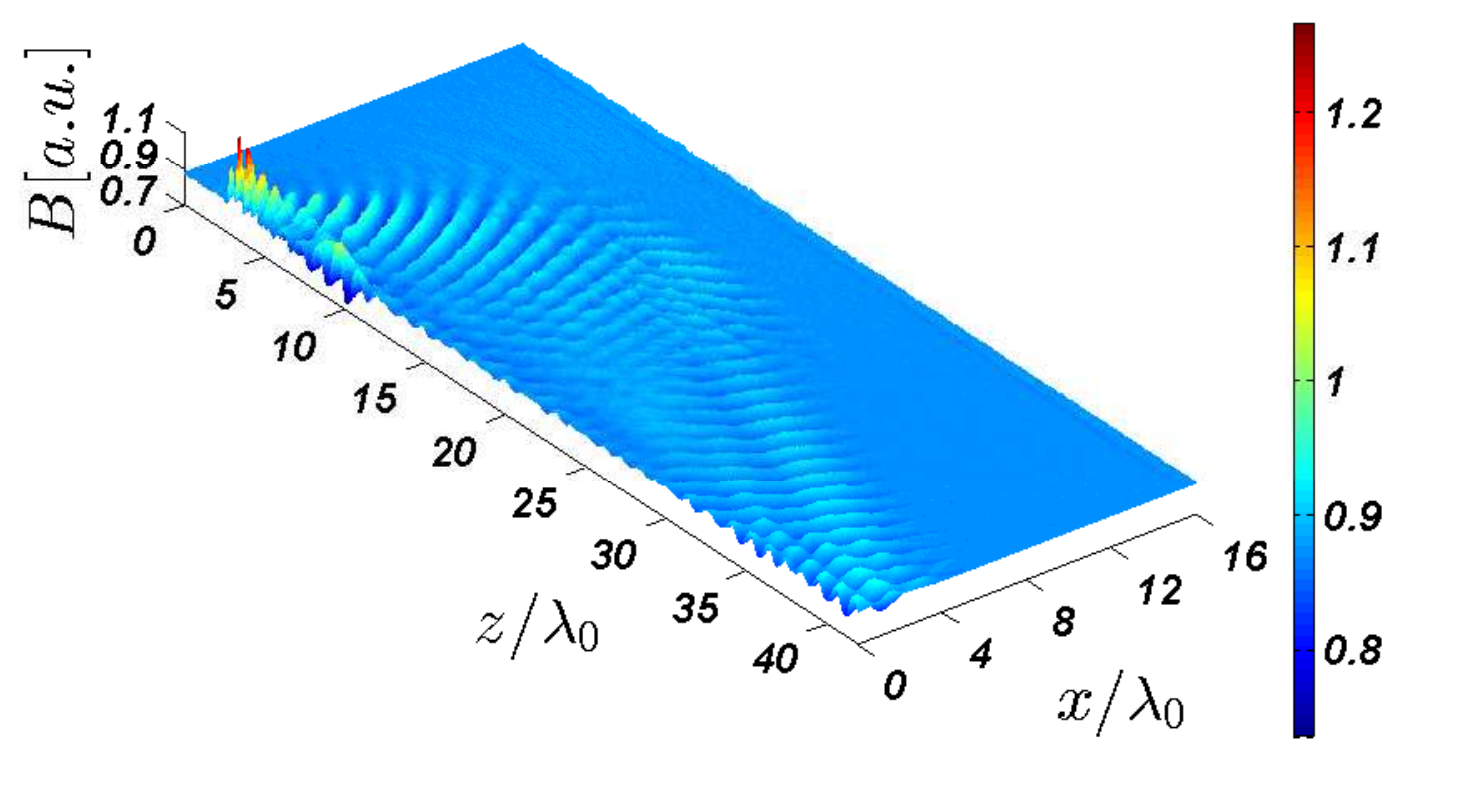}}\caption{Distribution of the magnitude of (a) electric and (b) magnetic fields
over the simulated region at time $t=50\, T_{wave}$, where $T_{wave}=1/f_{0}$
is the time period of the launched wave.\label{fig:EB}}
\end{figure*}

As shown in Figure \ref{fig:EB}, by considering the electric and
magnetic fields of the waves, we can easily observe the O-X conversion
near the O cut-off layer e.g. $x_{O-cutoff}=10.5\,\lambda_{0}$.

\begin{figure*}
\subfloat[$E_{x}$]{\includegraphics[scale=0.5]{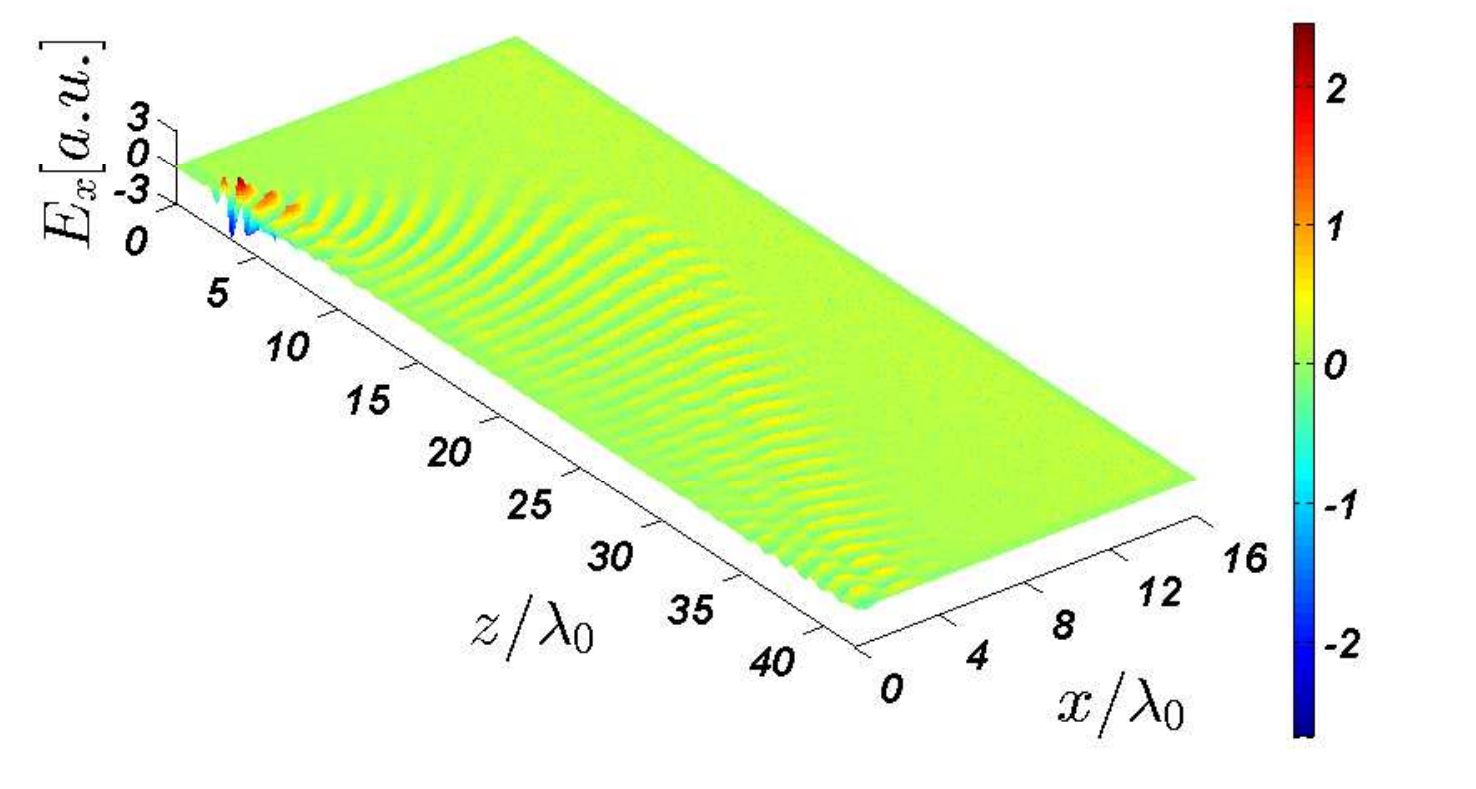}}\subfloat[$B_{x}$]{\includegraphics[scale=0.5]{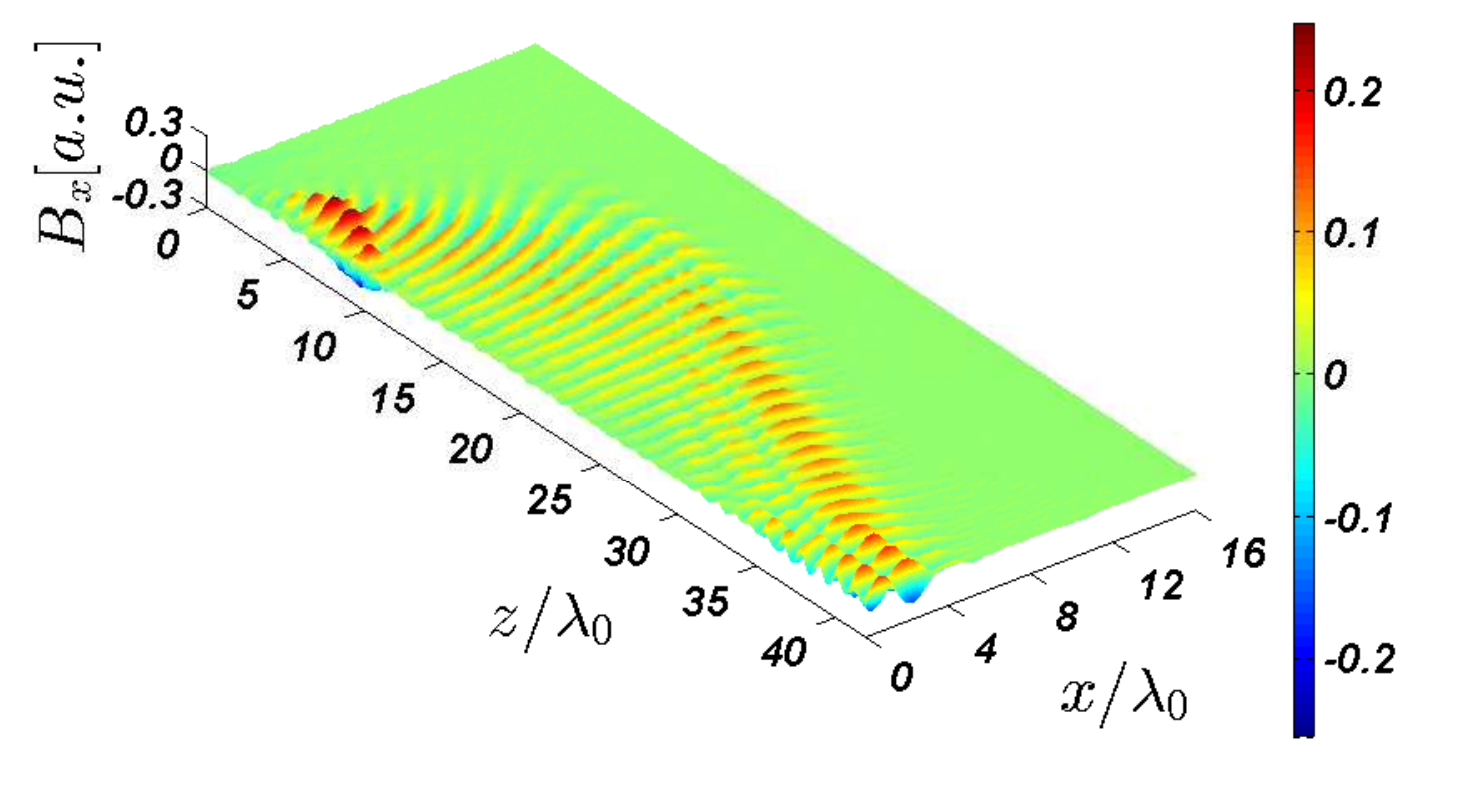}}\medskip{}
\subfloat[$E_{y}$]{\includegraphics[scale=0.5]{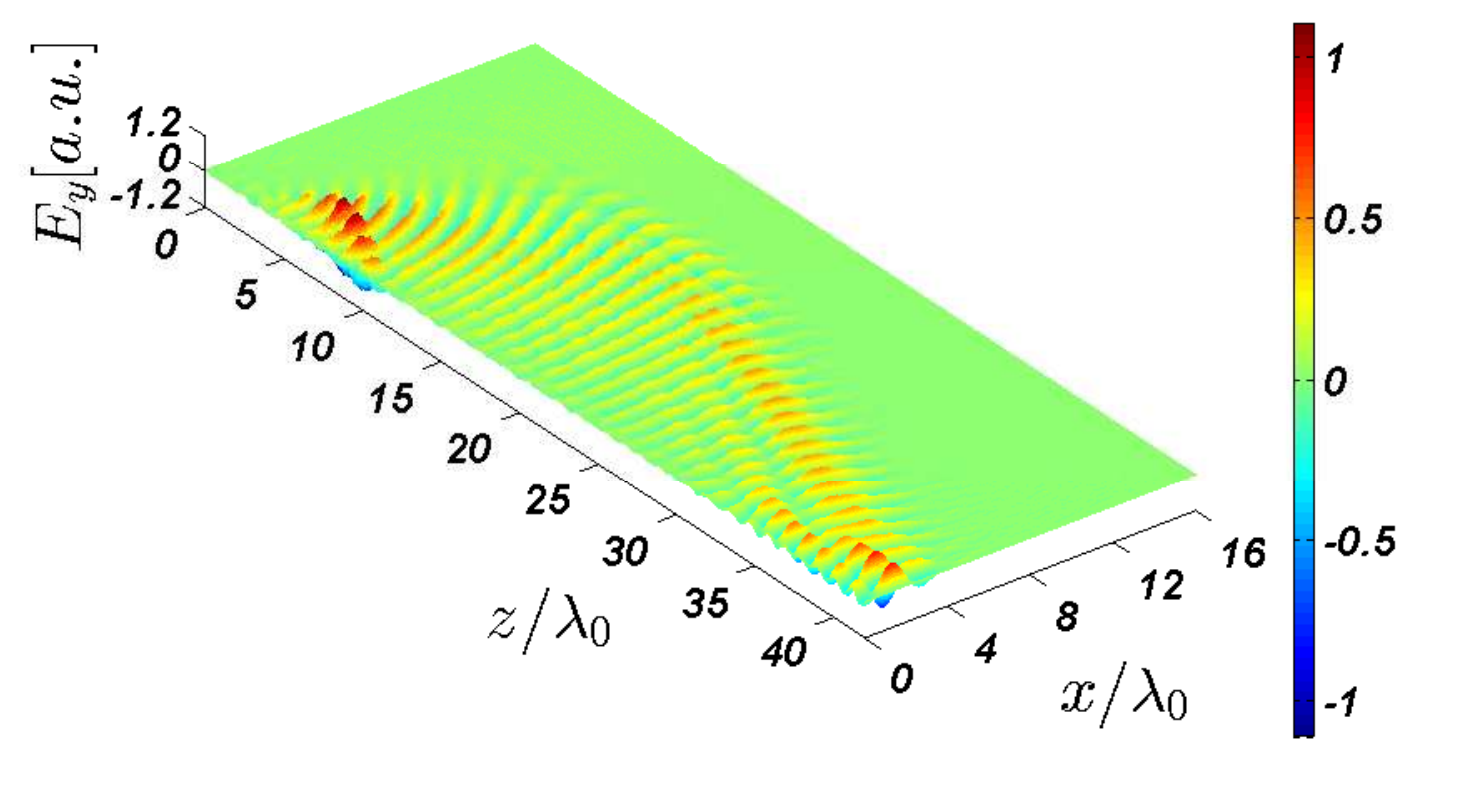}}\subfloat[$B_{y}$]{\includegraphics[scale=0.5]{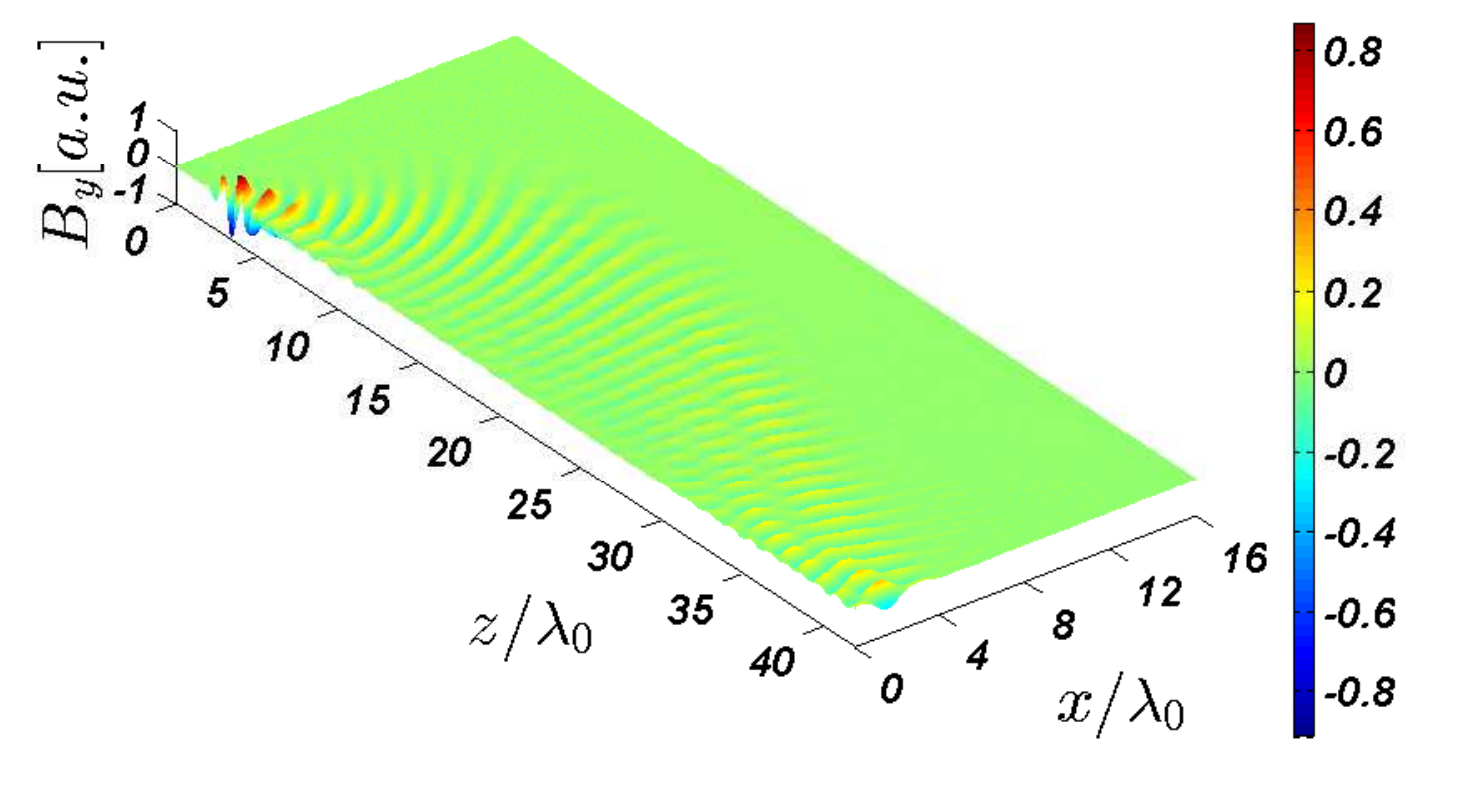}}\medskip{}
\subfloat[$E_{z}$]{\includegraphics[scale=0.5]{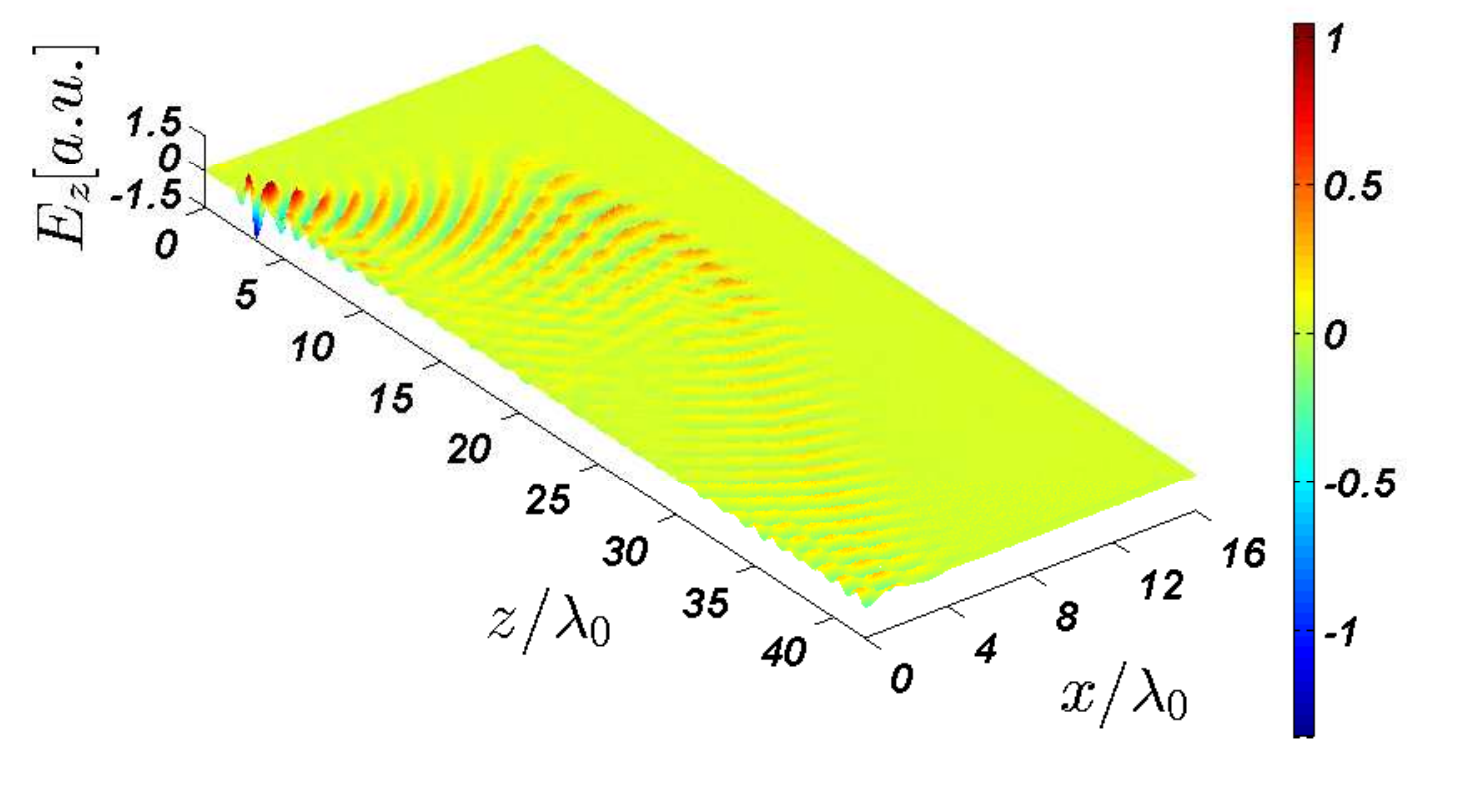}}\subfloat[$B_{z}$]{\includegraphics[scale=0.5]{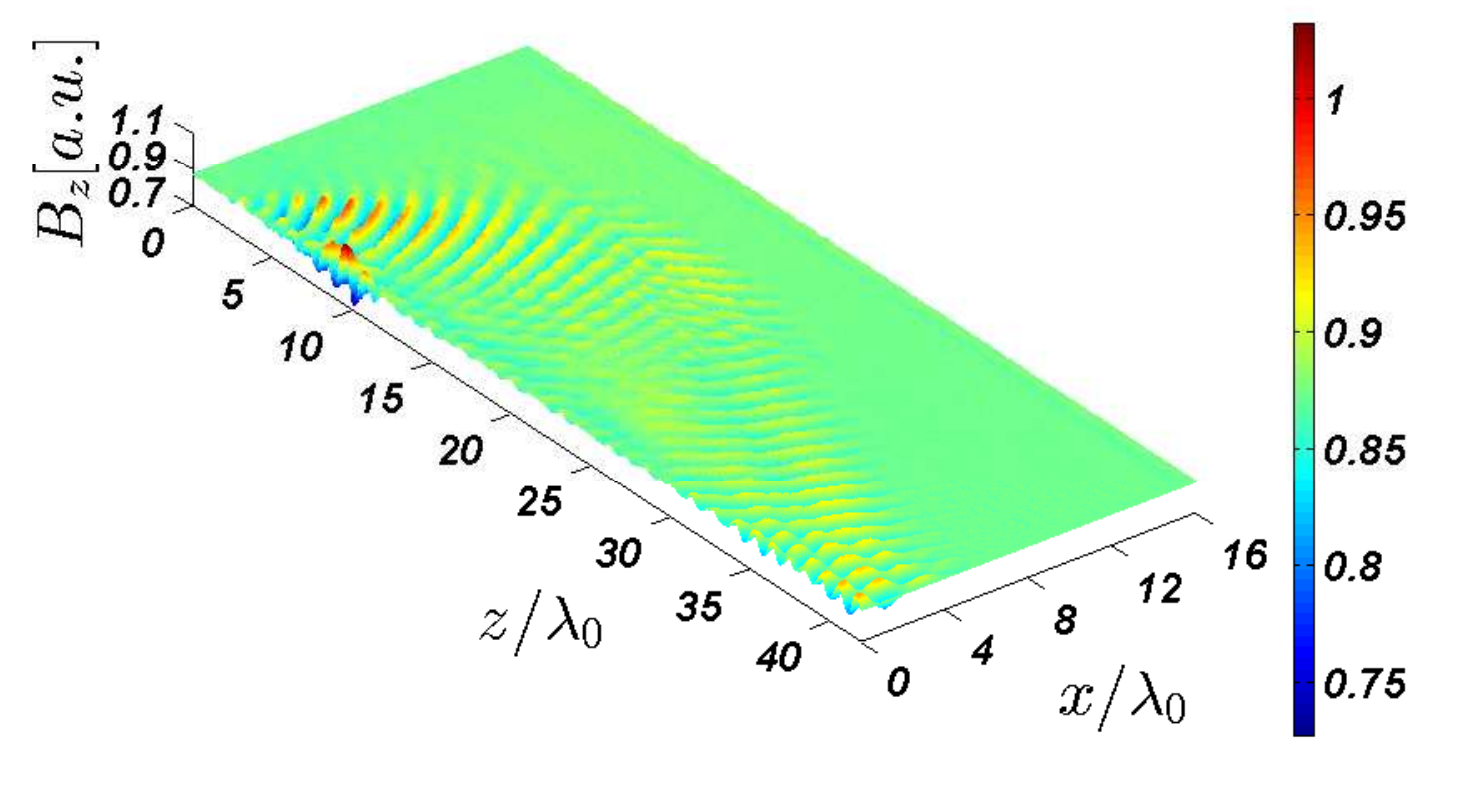}}\caption{Distribution of electric and magnetic field components over space
of the simulated region at time $t=50\, T_{wave}$, where $T_{wave}=1/f_{0}$
is the time period of the launched wave.\label{fig:EB components}}
\end{figure*}

Based on Figure \ref{fig:System simulation}, the reflected X-mode
wave should have the $E_{y}$ component of electric field and the
$B_{x}$ and $B_{z}$ components of magnetic field. By considering
Figure \ref{fig:EB components}, evidence of these waves can be seen.
For the reflected wave, $E_{y}$, $B_{x}$ and $B_{z}$ in comparison
with other components are stronger, indicating the presence of the
X-mode wave.

As is evident in Figures \ref{fig:EB} and \ref{fig:EB components},
we could not observe the X-B conversion near the UHR layer, $x_{UHR}=3.747\,\lambda_{0}$.
For this purpose we should increase number of grid cells and computational
particles in the upper hybrid resonance layer. To achieve this, we
must increase the number of computer processors to have a reasonable
run time. Also we should involve thermal effects and thermal gradient
in the simulation to prepare conditions for EBW propagation. In this
case, then we can consider other aspects of the O-X-B double conversion.
We can consider the influence of the density gradient on the conversion
efficiency, and obtain the optimum gradient. Also we can consider
influence of launch angle on the conversion efficiency and determine
the angle for maximum net conversion, and finally we can consider
parametric instability in the X-B conversion \cite{parametric instability}
and specify the power threshold of the instability.

\subsection{O-X conversion efficiency}

\begin{figure*}
\noindent \begin{centering}
\includegraphics[scale=0.9]{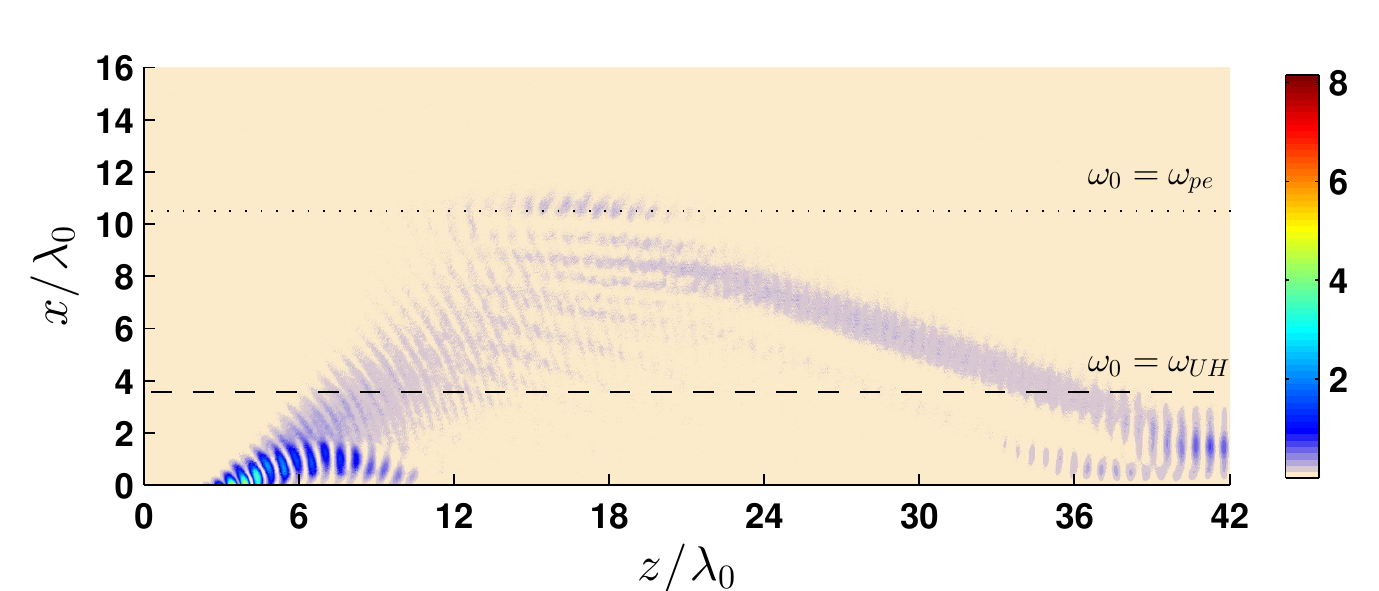}
\par\end{centering}

\caption{Distribution of electromagnetic power over the simulated region in
arbitrary units at time $t=50\, T_{wave}$, where $T_{wave}=1/f_{0}$
is the time period of the launched wave. The UHR and cutoff layers
are indicated by dashed and dotted lines, respectively.\label{fig:power2d}}
\end{figure*}

For considering the dependence of O-X conversion efficiency on density
inhomogeneity and the angle of the launched wave, first we calculate
this efficiency. We use the ratio of power propagating in the O-mode
wave to the conversion region to the power transported by the X-mode
wave out of this region. As shown in Figure \ref{fig:power2d}, we
can observe the process of O-X conversion by following the path of
power propagation. Power flow near the UHR layer will ensure inclusion
of incoming and outgoing power to the conversion region in our computations. 

\begin{figure*}
\subfloat[\label{fig:power1da}]{\noindent \centering{}\includegraphics{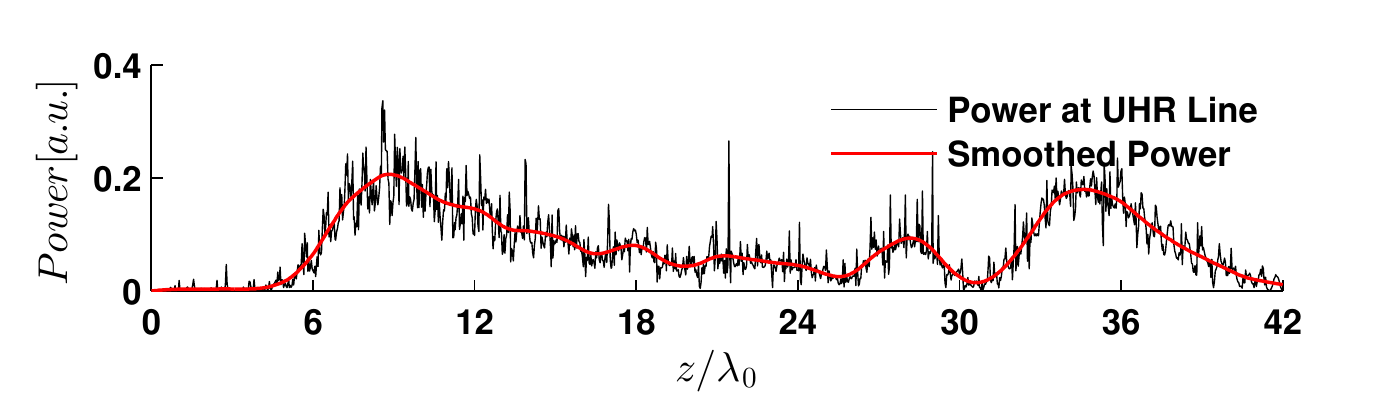}}\medskip{}
\subfloat[\label{fig:power1db}]{\noindent \centering{}\includegraphics{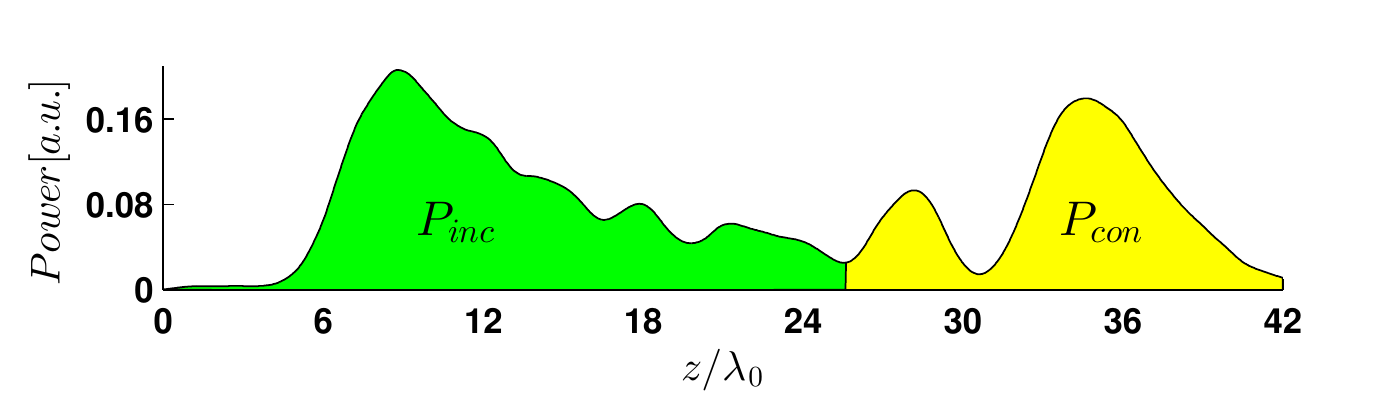}}\caption{(a) The noisy and smoothed power along the UHR line at time $t=50\, T_{wave}$,
where $T_{wave}=1/f_{0}$ is the time period of the launched wave.
(b) Based on Figure \ref{fig:power2d}, the regions related to the
launched O-mode wave and reflected X-mode wave are determined for
the smoothed power and filled up with green and yellow color, respectively.
The area under the curves is a measure of incident and converted power.\label{fig:power1d}}
\end{figure*}

The power of the electromagnetic fields along the UHR line is shown
in Figure \ref{fig:power1d}. Because of noisy nature of PIC simulation,
this diagram exhibits a high fluctuation level, so as shown in Figure
\ref{fig:power1da}, we fit the noisy diagram with a smoothing spline
function by applying the De Boor approach \cite{Smoothing spline,De Boor approach}.
For the noisy power $P(z_{i})$, the smoothed power $\tilde{P}(z)$
has been defined in such a way to minimize
\[
\alpha\sum_{i=1}^{n_{z}}\left[\frac{P(z_{i})-\tilde{P}(z_{i})}{\delta_{i}}\right]^{2}+(1-\alpha)\intop_{0}^{L_{z}}\left[\frac{d^{2}\tilde{P}(z)}{dz^{2}}\right]^{2}dz
\]
where $\alpha=0.52$ is the specified smoothing parameter and $\delta_{i}=1$
are the specified weights. The area under the curve of the incoming
O-mode wave is a measure of incident power $P_{inc}$ flowing into
the conversion region, and the area under the curve of the outgoing
X-mode wave is a measure of converted power $P_{con}$ after mode
conversion. These areas are shown in Figure \ref{fig:power1db}. With
these parameters we can define conversion efficiency as $e_{con}=P_{con}/P_{inc}$.
The parameters of this simulation are the same as used in the full
wave model of \cite{TJ-II kohn simulation}, and the obtained efficiency
$e_{con}=0.66$ for $4\,\lambda_{0}$ opening width and $\theta_{opt}=47^{\circ}$,
agrees well with this case in Table 1 in \cite{TJ-II kohn simulation}.

\section{Summary and Conclusion}

A kinetic particle model for O-X conversion in a dense magnetized
plasma is developed, and applied to O-mode launched at angle $\theta_{opt}=47^{\circ}$
for TJ-II parameters. The results, in good agreement with a full wave
model, obtained 66\% for conversion efficiency. The advantage of the
kinetic model in comparison with other models like the full wave model
is the ability to simulate the processes including collisional and
cyclotron damping which have important effects on X-B conversion and
EBW propagation. These dampings have a major impact on X-B conversion
because the X-mode wave during a collisional damping converts to EBW
and EBW after propagation in high density regions during a cyclotron
damping transfers wave energy to plasma and leads to plasma heating
\cite{full wave model}. 

For future work, we can consider the effects of the power of the launched
wave on conversion efficiency, and also on instabilities such as the
parametric instability. Effects of the density and external magnetic
field profiles on conversion and instabilities will also be studied.
Access to higher performance computing power will also make it possible
to resolve the X-wave to EBW conversion process, which requires resolution
of the EBW wavelength of the order of the electron gyro radius and
sufficient computer particle density to resolve a longitudinal wave.

\end{document}